\documentclass[%
superscriptaddress,
preprint,
 amsmath,amssymb,
 aps,
prl,
]{revtex4-2}

\usepackage{graphicx}% Include figure files
\usepackage{dcolumn}% Align table columns on decimal point
\usepackage{bm}% bold math
\usepackage{color}

\begin{document}

%\preprint{APS/123-QED}

\title{Mid-Air Single-Sided Acoustic Levitation in High-Pressure Regions of Zero-Order Bessel Beams}% Force line breaks with \\

% =================================
% \author{
% Yusuke Koroyasu\(^{1}\),
% Christopher Stone\(^{2}\),
% Yoichi Ochiai\(^{3,4,5,6}\),
% Takayuki Hoshi\(^{3}\),
% Bruce W. Drinkwater\(^{2}\),
% Tatsuki Fushimi\(^{4,5,6}\)
% }
% \affiliation{\(^{1}\) Graduate School of Comprehensive Human Science, University of Tsukuba, Tsukuba, 305-8550, Ibaraki, Japan}
% \affiliation{\(^{2}\)School of Electrical, Electronic and Mechanical Engineering, University of Bristol, BS8 1TR Bristol, United Kingdom}
% \affiliation{\(^{3}\) Pixie Dust Technologies, Inc., Chuo-ku, 104-0028, Tokyo, Japan}
% \affiliation{\(^{4}\) R\&D Center for Digital Nature, University of Tsukuba, Tsukuba, 305-8550, Ibaraki, Japan}
% \affiliation{\(^{5}\) Institute of Library, Information and Media Science, University of Tsukuba, Tsukuba, 305-9550, Ibaraki, Japan}
% \affiliation{\(^{6}\) Tsukuba Institute for Advanced Research (TIAR), University of Tsukuba, Tsukuba, 305-8577, Ibaraki, Japan}

% \email{koroyu@digitalnature.slis.tsukuba.ac.jp} % 対応著者
% ==================================

\author{Yusuke Koroyasu}
\email{koroyu@digitalnature.slis.tsukuba.ac.jp} % 対応著者
\affiliation{Graduate School of Comprehensive Human Science, University of Tsukuba, Tsukuba, Ibaraki 305-8550, Japan}

\author{Christopher Stone}
\affiliation{School of Electrical, Electronic and Mechanical Engineering, University of Bristol, Bristol BS8 1TR, United Kingdom}

\author{Yoichi Ochiai}
\affiliation{Pixie Dust Technologies, Inc., Chuo-ku, Tokyo 104-0028, Japan}
\affiliation{R\&D Center for Digital Nature, University of Tsukuba, Tsukuba, Ibaraki 305-8550, Japan}
\affiliation{Institute of Library, Information and Media Science, University of Tsukuba, Tsukuba, Ibaraki 305-9550, Japan}
\affiliation{Tsukuba Institute for Advanced Research (TIAR), University of Tsukuba, Tsukuba, Ibaraki 305-8577, Japan}

\author{Takayuki Hoshi}
\affiliation{Pixie Dust Technologies, Inc., Chuo-ku, Tokyo 104-0028, Japan}

\author{Bruce W. Drinkwater}
\affiliation{School of Electrical, Electronic and Mechanical Engineering, University of Bristol, Bristol BS8 1TR, United Kingdom}

\author{Tatsuki Fushimi}
\affiliation{R\&D Center for Digital Nature, University of Tsukuba, Tsukuba, Ibaraki 305-8550, Japan}
\affiliation{Institute of Library, Information and Media Science, University of Tsukuba, Tsukuba, Ibaraki 305-9550, Japan}
\affiliation{Tsukuba Institute for Advanced Research (TIAR), University of Tsukuba, Tsukuba, Ibaraki 305-8577, Japan}

% ==================================
% \author{Yusuke Koroyasu}
%  \altaffiliation[Also at ]{Physics Department, XYZ University.}%Lines break automatically or can be forced with \\
%  \email{Second.Author@institution.edu}
% \affiliation{%
%  Authors' institution and/or address\\
%  This line break forced with \textbackslash\textbackslash
% }%

% \collaboration{MUSO Collaboration}%\noaffiliation

%  \homepage{http://www.Second.institution.edu/~Charlie.Author}
% \affiliation{
%  Second institution and/or address\\
%  This line break forced% with \\
% }%
% \affiliation{
%  Third institution, the second for Charlie Author
% \affiliation{%
%  Authors' institution and/or address\\
%  This line break forced with \textbackslash\textbackslash
% }%

% \collaboration{CLEO Collaboration}%\noaffiliation
% ====================================

\date{\today}% It is always \today, today,

%TC:ignore
\begin{abstract} % 593<600
Acoustic levitation enables non-contact manipulation using sound waves. While conventional methods entrap particles at pressure nodes (zero-pressure region surrounded by high-pressure), we demonstrate stable acoustic levitation and translation in mid-air within a high-pressure axial core of a single-sided zero-order Bessel beam for the first time. The trap operates at a long working distance, up to 397 mm ($46.6 \lambda$), supports simultaneous multi-particle levitation, and maintains stability over obstacles. Our work establishes a new paradigm for single-sided acoustic manipulation in mid-air.
\end{abstract}
%TC:endignore
% 前のversion [10/03] 70charくらい減らしたい。
% Acoustic levitation enables non-contact manipulation using sound waves, benefiting many scientific and engineering applications. While conventional methods entrap particles at pressure nodes (zero-pressure points surrounded by high-pressure regions), we demonstrate stable acoustic levitation and translation in mid-air within pressure antinodes (high-pressure axial core) of a single-sided Bessel beam for the first time. The trap operates at a long working distance, up to 397 mm ($\sim 46.6 \lambda$), supports simultaneous multi-particle levitation, and maintains stability over obstacles. Our work establishes a new paradigm for single-sided acoustic manipulation in mid-air.

\maketitle

%\tableofcontents

%\paragraph{Introduction.—}
% 日本語コメント：standingwave~twin trap。従来の低圧領域でのトラップ
\textit{Introduction—}Acoustic levitation has emerged as a key technique for non-contact manipulation, with applications spanning from laboratory automation~\cite{Foresti2013a, andrade2018automatic, brotton2020controlled, zang2017acoustic}, volumetric displays~\cite{fushimi2019acoustophoretic, hirayama2019volumetric, ochiai2014pixie} to additive manufacturing~\cite{ezcurdia2022leviprint, chen2025acoustics, chen2025omnidirectional, melde2023compact}. 
Most of these implementations rely on standing wave configurations, in which the behavior of subwavelength particles (Rayleigh regime) is governed by the acoustic contrast arising from differences in compressibility and density between the particle and the surrounding medium. 
In a standing wave in air, virtually all particles are consistently trapped at the pressure nodes surrounded by high pressure regions~\cite{bruus2012acoustofluidics}. Single-sided levitators implement the same principle by generating twin, vortex, and bottle traps that enclose a central minima with high-pressure regions~\cite{marzo2015holographic, baresch2016observation, choi2024magnetically, wu2025real}. 

% \begin{figure}[t]
%     \centering
%     % Mock content for the figure, replace with actual \includegraphics
%     \includegraphics[width=1.0\columnwidth]{figs/concept_diagram.pdf} % Example for a combined figure
%     \caption{Illustration of particle levitation on the central core of zero-order Bessel beam. The beam is generated using PATs.}
%     \label{fig:concept_diagram}
% \end{figure}

% 第2段落：課題提示と理論的解決策。最後に、分野の課題を要約する。「浮遊には正のコントラスト因子が必要」という従来のパラダイムと、実験的な困難さから、高圧領域での3D浮遊は実現不可能と見なされてきた、という背景を述べる。
While entrapment of Rayleigh-sized particles at pressure nodes is a general consensus in mid-air levitation, recent theoretical studies have predicted the possibility of trapping particles within high-pressure antinodes in mid-air. Fan and Zhang, for example, theoretically investigated transverse (2D) trapping in a zero-order Bessel beam, a field with a high-pressure axial core, predicting transverse stability is governed by an interplay between the beam's cone angle and particle-to-medium ratios of density and bulk modulus~\cite{fan2019trapping}. The predicted transverse restoring force was later experimentally demonstrated with a tethered sphere (i.e., not levitation)~\cite{fan2023transverse}. The pressure distribution for a zero-order Bessel beam propagating along a $z$-axis is given by~\cite{durnin1987diffraction, durnin1987exact}
\begin{equation}
p(r, z, t) = P_A J_0(kr\sin\beta)e^{j(kz\cos\beta-\omega t)},
\label{eq:BesselPressure}
\end{equation}
where $P_A$ is the amplitude, $J_0$ is the zero-order Bessel function, $k$ is the wavenumber, $\omega$ is the angular frequency, $\beta$ is the cone angle, and $r=\sqrt{x^2+y^2}$ is the radial coordinates, respectively. Ref.~\cite{fan2019trapping} showed that the 2D transverse stability is governed by a beam-parameter ($\beta$), predicting sets of condition for particles to be trapped at high-pressure regions in air. Axially, such high-pressure beams exert a propulsive force sufficient to counteract gravity. However, extending the transverse (two-dimensional) stability to achieve stable three-dimensional levitation has remained an experimental challenge. This difficulty is initially described by Marzo \textit{et al.}, who characterized levitation at an acoustic focus (high pressure region) as an experimentally unstable theoretical possibility~\cite{marzo2015holographic}, a finding that we were also able to replicate (Fig.~S3, S4 and Movie~S1 in Ref.~\cite{SM}). In general, although theoretical models allow for levitation at high-pressure regions, experiments to date show particles levitating only at low-pressure nodes.

\begin{figure*}[t]
    \centering
    \includegraphics[width=0.95\textwidth]{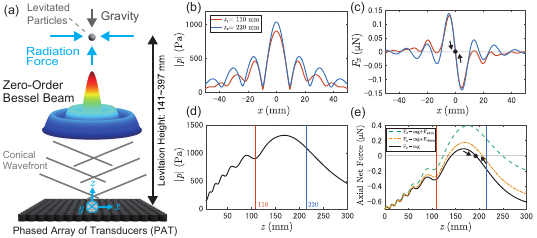}
    \caption{
        (a) Illustration of particle levitation at the central core of a zero-order Bessel beam. The beam is generated using a PAT. (b)–(e) Calculated acoustic pressure field of a zero-order Bessel beam from the PAT ($\beta = 20^\circ$, $V_\text{in}=8\,\text{V}$) and the radiation force on an EPS sphere (radius $a=0.75\,\text{mm}$). (b) Transverse and (d) axial pressure profiles. (c) Transverse force $F_x$. For the axial plots (d) and (e), solid vertical lines indicate the two evaluation planes, \(z_1=110\,\text{mm}\) (red) and \(z_2=220\,\text{mm}\) (blue, near the experimental levitation height). (e) Net axial force. The black line shows the balance between radiation force and gravity. The orange and green dashed lines include the drag from acoustic streaming calculated with thermoviscous~\cite{hu2015sound} and atmospheric~\cite{bass1995atmospheric} attenuation models, respectively. The range between these two models is consistent with experimentally observed streaming effects~\cite{stone2025characterising, stone2025experimental}.
    }
    \label{fig:ForcesBeta20}
\end{figure*}

% 日本語コメント：第四段落では、本研究のブレークスルーを宣言し、その物理的基盤、片側浮遊における新規モードとしての位置づけ、そして従来法を凌駕するユニークな利点（自己修復性、そして数値で示す長作動距離）を具体的に述べ、本研究の意義で締めくくる。
In this Letter, for the first time, we experimentally demonstrate three-dimensional acoustic levitation within the high-pressure axial core of a zero-order Bessel beam in mid-air [Fig.~\ref{fig:ForcesBeta20}(a)]. The trap is formed by a transverse restoring force~\cite{fan2019trapping,fan2023transverse} and an axial radiation force~\cite{fan2019trapping,marston2006axial,marston2007scattering} that counteracts gravity. This result establishes a novel mode of acoustic levitation distinct from the entrapment in planar standing waves and conventional single-sided twin or vortex traps. It supports real-time three-dimensional position control, simultaneous levitation of multiple particles, and the levitation of non-spherical objects. The levitation capability remains intact through a physical obstruction due to the Bessel beam's diffraction-free and self-healing ability ~\cite{garces2002simultaneous, antonacci2019demonstration}. Furthermore, the axial high-pressure remains confined, extending the working distance from 67 mm ($7.9\lambda$) in existing single-side traps to 397 mm ($46.6\lambda$) with the Bessel beam. These findings open new pathways for long-range acoustic manipulation in an open environment, free from the enclosures of standing wave configurations.

% ----------------------------Method-------------------------------
% 日本語コメント：この段落では、実験装置とビーム生成手法、そして実験対象について記述します。まず、超音波フェーズドアレイを用いたゼロ次ベッセルビームの生成が既存技術であることを述べ、引用を追加します。次に、本研究で使用したアレイの仕様と、ビームを電子的に合成・走査するための位相計算式を提示します。最後に、主な実験対象である粒子の物性値とその測定方法を定義します。
\textit{Methods—}The zero-order Bessel beam is generated using a $16 \times 16$ phased array of transducers (PAT, Murata MA40S4S, $40\,\text{kHz}$) known as SonicSurface~\cite{morales2021generating}, as shown in Fig~\ref{fig:ForcesBeta20}(a). The phase $\phi_i$ of each transducer at position $(x_i, y_i, z_i)$ is given by~\cite{hasegawa2017electronically, norasikin2019sonicspray}:
\begin{equation}
\phi_i = -k( \sin\beta\sqrt{x_i^2+y_i^2} - \cos\beta z_i).
\label{eq:bessel_phase}
\end{equation}
The beam was electronically steered by applying a virtual rotation to the transducer coordinates~\cite{hasegawa2017electronically}. Acoustic pressure field measurements verified that the generated beam profile corresponds well with numerical predictions (Fig.~S5 in Ref.~\cite{SM}). We used an expanded polystyrene (EPS) sphere of radius $a=0.75\,\text{mm}$ ($\frac{a}{\lambda}=0.088$), similar in scale to particles commonly employed in previous mid-air levitation experiments~\cite{marzo2015holographic, marzo2019holographic, hirayama2019volumetric, fushimi2019acoustophoretic}. The particle density was $\rho_p=40.4\,\text{kg\,m}^{-3}$ and unless otherwise stated, this particle was used throughout the experiments. 

\begin{figure*}[t]
    \centering
    % Mock content for the figure, replace with actual \includegraphics
    \includegraphics[width=0.95\textwidth]{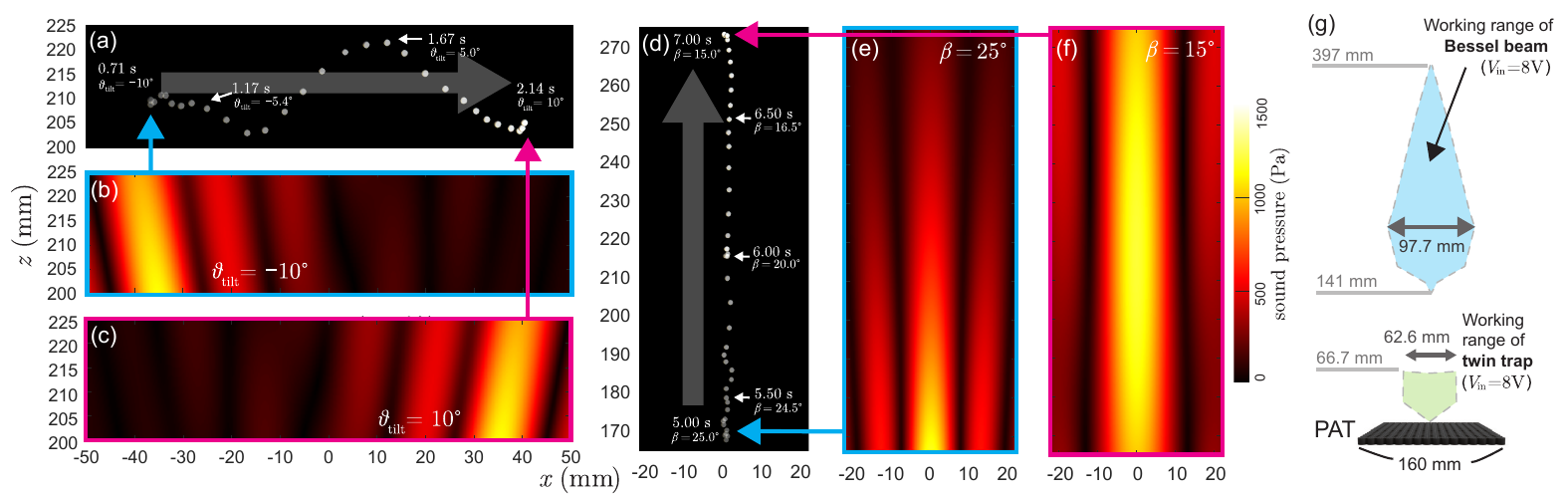} % Example for a combined figure
\caption{
Particle translation by (a)-(c) horizontal beam tilting ($\theta_\text{tilt}$ from -10$^{\circ}$ to 10$^{\circ}$) and (d)-(f) vertical cone angle modulation ($\beta$ from 25$^{\circ}$ to 15$^{\circ}$). The experimental trajectories in (a) and (d) are composites of images taken every 0.05 s, visualized with a 30--100\% transparency ramp. The corresponding simulated pressure profiles are shown in (b, c) and (e, f). (g) Experimental comparison of the working ranges for a Bessel beam and a conventional twin trap (average of 5 trials).
}
    \label{fig:manipulate}
\end{figure*}

% -------------------------数値計算--------------------------
%【段落の役割】数値計算により、ゼロ次ベッセルビームが軸上の高圧領域に安定な三次元トラップを形成する物理的根拠を提示する。
\textit{Results—}To evaluate the levitation capability, the acoustic radiation force on an EPS sphere was numerically calculated using Andersson and Ahrens method, which utilizes spherical harmonic expansion of the PAT-generated zero-order Bessel beam ($\beta=20^\circ$) with compressible-sphere scattering coefficients~\cite{Andersson2019, Andersson2022, zehnter2021acoustic, SM}. The calculations were performed at two axial positions chosen to bracket the on-axis pressure maximum, \(z_1 = 110\,\text{mm}\) ($12.9\lambda$) and \(z_2 = 220\,\text{mm}\) ($25.8\lambda$). As shown in Figs.~\ref{fig:ForcesBeta20}(b) and \ref{fig:ForcesBeta20}(c), the transverse force $F_x$ at both positions is restoring, directed toward the beam axis ($x=0$) despite an on-axis pressure maximum. This restoring force provides lateral confinement over ranges of approximately $\pm 12.1\,\text{mm}$ ($1.42\lambda$) at $z_1$ and $\pm 10.2\,\text{mm}$ ($1.20\lambda$) at $z_2$. Figure~\ref{fig:ForcesBeta20}(d) plots the axial pressure profile, and Fig.~\ref{fig:ForcesBeta20}(e) plots the corresponding axial net force, which indicates an equilibrium at $z \sim 192~\text{mm}$ ($22.5\lambda$) where the upward acoustic radiation force balances gravity. The effect of acoustic streaming ($F_\text{atm}$ and $F_\text{therm}$) will be discussed later. This axial balance between the upward radiation force and gravity, combined with the lateral confinement, forms a full three-dimensional trap. The axial restoring force remains directed toward the equilibrium for positions $z \gtrsim 141~\text{mm}$ ($16.5\lambda$). A detailed force profile in the $xz$-plane is provided in Fig.~S6 of Ref.~\cite{SM}. At the numerical equilibrium point ($z=192~\text{mm}$), the axial stiffness, $6.57~\mu\text{N m}^{-1}$, is an order of magnitude lower than the lateral stiffness, $70.4~\mu\text{N m}^{-1}$, similarly to single-sided twin and vortex traps~\cite{marzo2015holographic}.

% ========================実験=============================
%日本語コメント：この段落では、前段落で提示した数値計算による予測を実験的に検証します。まず、全ての試行において安定浮遊が成功したことを示し、トラップの堅牢性を確立します。次に、観測された粒子の軸方向および横方向の挙動を、平均値と標準偏差を含む具体的な数値で定量的に記述します。その後、放射力のみの理論予測との差異を明確に指摘し、その差異が音響流に起因することで説明可能であることを示します。先行研究（Stone et al.）が実験的な音響流「速度」は熱粘性減衰と大気減衰モデルの予測の間に収まることを示した事実を正確に引用し、この知見に基づいて我々の系で2つの減衰モデルから抗力を見積もり、計算された平衡高さの範囲を提示します。最後に、観測された浮上高さがこの範囲内に収まることから、音響流が軸方向の力のバランスに寄与していることを示唆して、この段落の報告を終えます。
To experimentally validate the trapping capability, we conducted 15 independent trials using a zero-order Bessel beam with $\beta = 20^{\circ}$ (see Movie~S2, Fig.~S10 in Ref.~\cite{SM}). In each trial, a particle was stably levitated for the entire 60 s observation period. During these trials, the particle's axial coordinate fluctuated between $205.6\pm 8.2\,\text{mm}$ and $233.5\pm 2.1\,\text{mm}$, with a mean levitation height of $220.4\pm 0.8\,\text{mm}$. Simultaneously, it exhibited transverse fluctuations over a range of $5.8\pm 0.9\,\text{mm}$. The particle also exhibited spontaneous transitions among discrete equilibrium heights within its axial range (see details in later discussions). 

The levitation capability of the Bessel beam is not limited to a spherical object, and levitation can also be performed with an non-spherical dried tea leaf (approximate diameter, 0.8 mm; $V_\text{in}=16\,\text{V}$; Fig.~S11(a) and Movie S3 in Ref.~\cite{SM}), a flat piece of silica aerogel (Classic~Silica\textsuperscript{TM} Disc, Aerogel Technologies, LLC; major axis, ~2.3 mm; $V_\text{in}=8\,\text{V}$; Fig.~S11(b) and Movie~S4 in Ref.~\cite{SM}), and a single disk formed from coalesced potato starch powder (approximate diameter, 1.3 mm; $V_\text{in}=15\,\text{V}$; Movie~S5 in Ref.~\cite{SM}).

% 日本語コメント：この段落では、ビームの傾斜と円錐角の変調が粒子の運動に与える影響を調査し、ビーム走査による粒子操作の可能性を実証します。
We further explored the particle translation capabilities of the Bessel beam by modulating the beam property [Figs.~\ref{fig:manipulate}(a)-\ref{fig:manipulate}(f)]. Tilting the Bessel beam generates horizontal forces directed toward the high-intensity central axis of the beam, thereby enabling the horizontal translation of particles. Specifically, the tilt angle was modulated according to \(\theta_{\text{tilt}}(t) = A_{\text{tilt}} \sin(2\pi f_{\text{tilt}} t+\pi)\), where $t$ represents time, \(A_{\text{tilt}} = 10^\circ\) is the maximum tilt angle, and \(f_{\text{tilt}} = 0.35\) Hz is the modulation frequency. Using these parameters, the particles were translated horizontally at a speed of approximately 5.7 cm s\(^{-1}\) (Figs.~\ref{fig:manipulate}(a)-\ref{fig:manipulate}(c) and Movie~S6 in Ref.~\cite{SM}). The particle was successfully translated horizontally in response to the modulated tilt angle of the beam. This modulation also introduced slight vertical oscillations on the order of centimeters (Figs.~S12(a) and S12(b) in Ref.~\cite{SM}).

Two methods were considered for the vertical translation: modulation of the cone angle (\(\beta\)) of the Bessel beam and modulation of the input voltage ($V_{\text{in}}$) to the PAT. We employed the cone angle modulation method over the voltage modulation method because of its higher stability in preliminary experiments. This control is achieved through the monotonic dependence of the axial radiation force on the cone angle $\beta$. Decreasing $\beta$ enhances the upward force and raises the particle’s equilibrium position, whereas increasing $\beta$ reduces the force and lowers it (Eq. (12) in Ref.~\cite{fan2019trapping}). The cone angle was modulated according to \(\beta(t) = A_\beta \sin(2\pi f_\beta t+\pi) + \beta_{{\text{offset}}}\), where \(A_\beta = 5^\circ\) and \(f_\beta = 0. 25\) Hz are the amplitude and frequency of the cone angle modulation, respectively, and \(\beta_{{\text{offset}}} = 20^\circ\) is the offset cone angle. Using these parameters, the particles were translated vertically at a speed of approximately 4.3 cm s\(^{-1}\) (Figs.~\ref{fig:manipulate}(d)-\ref{fig:manipulate}(f), Figs.~S12(c) and S12(d), and Movie~S7 in Ref.~\cite{SM}).

% 日本語コメント：この段落では、水平・垂直方向の並進操作を組み合わせて実験的に決定されたベッセルビームの操作可能範囲（working range）を報告し、従来のツイン・トラップの範囲と比較することで、本手法の優位性を明確に示します。
By combining the horizontal and vertical translation techniques described above, we experimentally determined the working range achievable with the Bessel beam (see Methods in Ref.~\cite{SM}). As shown in Fig.~\ref{fig:manipulate}(g), the particles were translated over a horizontal distance of 97.7 mm ($11.5\lambda$) and a vertical range of 141–397 mm (16.5-46.6$\lambda$). For comparison, a conventional twin trap under identical conditions was limited to a horizontal range of 62.6 mm ($7.34\lambda$) and a maximum height of 66.7 mm ($7.82\lambda$), highlighting the significantly extended working range of our approach. 

\begin{figure}[t]
    \centering
    % Mock content for the figure, replace with actual \includegraphics
    \includegraphics[width=0.48\textwidth]{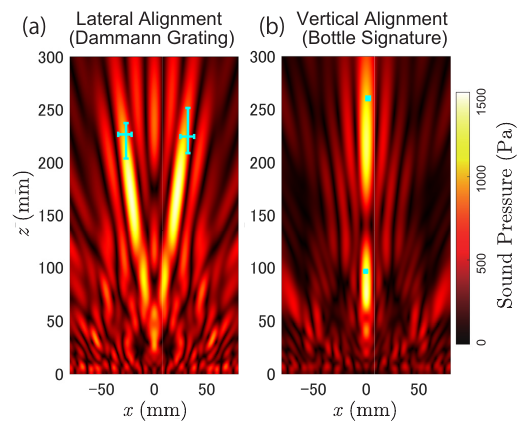} % Example for a combined figure
    \caption{Simultaneous levitation of multiple particles. (a) Parallel trapping of two particles (left: $a=0.79\,\text{mm}$, $\rho_p=32.3\,\text{kg\,m}^{-3}$; right: $a=0.79\,\text{mm}$, $\rho_p=31.6\,\text{kg\,m}^{-3}$) in replicated Bessel beams with Damman grating. (b) Vertically aligned trapping of two particles (top: $a=0.77\,\text{mm}$, $\rho_p=35.5\,\text{kg\,m}^{-3}$; bottom: $a=0.77\,\text{mm}$, $\rho_p=34.3\,\text{kg\,m}^{-3}$) in separated zones with bottle signature . The cyan lines indicate the range of positional fluctuations, which is significantly smaller for the vertically aligned trap.}
    \label{fig:other}
\end{figure}

% 日本語コメント：この段落では、次に、複数のベッセルビームを水平方向に生成し、複数の粒子を横方向に整列させて並行トラップする手法を実証します。
We further demonstrate the parallel trapping of multiple particles by generating multiple Bessel beams from a single PAT. This was achieved using an acoustic Dammann grating, a binary phase hologram used for beam replication in both optics~\cite{dammann1971high, vasara1989realization, Garcia-Martinez:12} and, more recently in acoustics~\cite{fushimi2024multi}. The grating inverts the phase by $\pi$ for transducers where $0.1A < |x_{tr}| < 0.3A$ ($A=160\,\text{mm}$ is the array size, Fig.~S2(a) in Ref.~\cite{SM}), which creates two distinct beams along the $x$-axis. With a cone angle of $\beta=20^\circ$ and an input voltage of $V_\text{in}=13\,\text{V}$, two particles were simultaneously levitated in the cores of the beams (Fig.~\ref{fig:other}(a) and Movie~S8 in Ref.~\cite{SM}). The two particles were trapped at time-averaged (20~s) equilibrium positions of $(x,z)=(-26.8, 227)\,\text{mm}$ (left) and $(32.3, 225)\,\text{mm}$ (right). Their respective positional fluctuations spanned $12.6$ and $13.0\,\text{mm}$ in the $x$-direction, and $33.6$ and $42.7\,\text{mm}$ in the $z$-direction.

% 日本語コメント：この段落では、次に、ボトルトラップ構成を用いて、複数の浮遊粒子を縦方向に整列させてトラップする手法を実証します。
A vertical alignment of levitated particles can be demonstrated using a bottle trap configuration. This approach splits the high-pressure core along the beam axis into two distinct trapping zones by creating a central low-pressure region. The bottle phase signature was implemented by inverting the phase by $\pi$ for transducers where $\sqrt{x_{i}^2+y_{i}^2} > 0.35A$ (Fig.~S2(b) in Ref.~\cite{SM} and Ref.~\cite{marzo2015holographic}). With an input voltage of $V_\text{in}=9\,\text{V}$ and a cone angle of $\beta=21^\circ$, two particles were simultaneously levitated in these vertically separated zones (Fig.~\ref{fig:other}(b) and Movie~S9 in Ref.~\cite{SM}). Their time-averaged (20~s) equilibrium positions were $(x,z)=(-0.588, 96.7)\,\text{mm}$ (bottom) and $(1.59, 261)\,\text{mm}$ (top), with respective positional fluctuations spanning $1.84$ and $3.33\,\text{mm}$ in the $x$-direction, and $1.72$ and $3.29\,\text{mm}$ in the $z$-direction.

\begin{figure}[b]
    \centering
    % Mock content for the figure, replace with actual \includegraphics
    \includegraphics[width=0.48\textwidth]{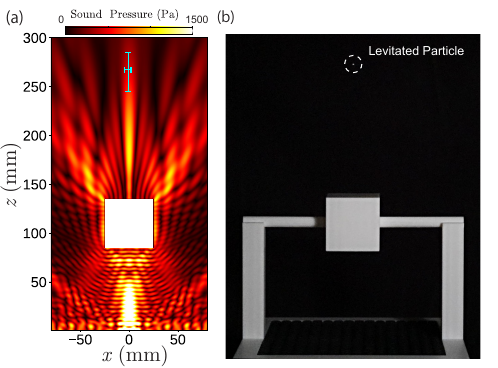} % Example for a combined figure
    \caption{Experimental demonstration and numerical simulation of levitation through an obstacle. (a) Simulated sound pressure field using BEM, showing the reconstruction of the Bessel beam profile beyond the obstacle. The mesh size for BEM was set to approximately $\frac{\lambda}{11}$. The cyan lines indicate the range of positional fluctuations. (b) Photograph of a particle levitated above the obstacle.}
    \label{fig:obstacle}
\end{figure}

% 日本語コメント：この段落では、ベッセルビームの自己修復特性を提示し、まずBEMによる数値計算でその現象を確認、次に実験で検証することで、他の手法に頼らない堅牢な操作が可能であることを示します。
We further demonstrate acoustic levitation through a physical obstruction, an application that builds on the diffraction-free and self-healing nature of a Bessel beam that enables them to maintain their pressure profile after an obstacle~\cite{antonacci2019demonstration}. Boundary element method (BEM) simulations \cite{hirayama2022high} with the AcousTools library~\cite{Mukherjee_AcousTools} show the clear reconstruction of the axial high-pressure core for a $\beta=20^\circ$ zero-order Bessel beam obstructed by a $50\,\text{mm}$ cube at $z_1=110\,\text{mm}$ [Fig.~\ref{fig:obstacle}(a)]. This prediction was confirmed experimentally: with an input voltage of $V_\text{in}=14\,\text{V}$, a particle was successfully levitated in the reconstructed beam (Fig.~\ref{fig:obstacle}(b) and Movie~S10 in Ref.~\cite{SM}). The particle was maintained at a time-averaged (20 s) equilibrium position of $(x,z)=(-0.652, 267)\,\text{mm}$, with positional fluctuations spanning $6.61\,\text{mm}$ in the $x$-direction and $40.2\,\text{mm}$ in the $z$-direction. 

% ============================================================
%\paragraph{Discussions and conclusions.—}
% 日本語コメント：【本研究の核心的成果、意義、利点の要約】この段落では、まず本研究の最も重要な貢献（高圧領域での3D浮遊の実証）を明確に宣言します。次に、その成果が従来の低圧領域トラップという常識を覆すものであることを示し、学術的な意義付けを行います。最後に、この新原理がもたらす主要な利点（長距離操作、自己修復性、開放環境での操作）を要約し、本研究のインパクトの全体像を提示します。
\textit{Discussion—}So far, we have experimentally demonstrated stable three-dimensional acoustic levitation within the high-pressure axial core of a zero-order Bessel beam. Though, numerical simulation generally points towards the experimentally observed results, we further explored the contribution of the various forces. 

% 日本語コメント：【軸方向の力の議論の簡潔化】この段落では、理論と実験の差異、その原因としての音響流、そしてモデル計算による裏付けという論理の流れを、より簡潔かつ直接的に表現します。手法の繰り返しを避け、物理的な議論の核心に焦点を絞ることで、冗長性を排除しています。
Firstly, the radiation force model alone predicts a levitation height of $z=192$ mm, whereas particles were observed to levitate at $220.4 \pm 0.8\,\text{mm}$. This difference could be improved by accounting for the upward drag from acoustic streaming. Our numerical simulations predict acoustic streaming velocities between $0.198\,\text{m\,s}^{-1}$ (thermoviscous model) to $0.493\,\text{m\,s}^{-1}$ (atmospheric model), depending on assumed attenuation model at the experimental levitation height ($z_2 = 220\,\text{mm}$). Indeed, our simulations incorporating this streaming effect predict an equilibrium range of $207$--$249\,\text{mm}$, in agreement with our experimental observations (Fig.~\ref{fig:ForcesBeta20}(e) and Fig.~S9 and Methods in Ref.~\cite{SM}). Thus, acoustic streaming is considered to be non-negligible contributor to the final particle position in axial direction.

% 日本語コメント：【復元力の物理的起源の説明】この段落では、本研究で観測された横方向の復元力の物理的起源を、Gor'kovポテンシャルに基づいて説明します。まず、復元力が特定の条件下（コーンアングルが閾値以下）でのみ生じることを述べ、その条件下で速度勾配項が圧力勾配項を上回ることが力の源泉であることを明らかにします。最後に、集束ビームにおける条件を客観的に併記することで、ベッセルビームの状況を相対化します。
As predicted by Gor'kov potential~\cite{gorkov1961dokl, SM}, a restoring force is generated in the transverse direction when the cone angle $\beta$ is below a critical threshold. For an ideal beam with an infinite aperture, this threshold is calculated to be $\beta < 26.8^\circ$ for an EPS sphere in air (Eq.~(9) in Ref.~\cite{fan2019trapping}). Under this condition, a dominant velocity-gradient term overcomes the competing pressure-gradient term, enabling confinement along the beam axis (Fig.~S13 in Ref.~\cite{SM}). For a conventional focused beam, this condition is met only for long focal lengths (e.g., $>$129~mm).

% 日本語コメント：【トラップ安定性のパラドックスと未解明な点】この段落では、実験的に観測されたBesselビームの予期せぬ安定性が、既存の放射力計算の結果と矛盾するという「パラドックス」を提示します。既知の二次的な力（音響流、回転力）ではこの矛盾を説明できないことを根拠と共に示した上で、この安定性の背後には未解明の物理的メカニズムが存在することを示唆し、本研究の新規性と重要性を強調します。
We also note a discrepancy between the predicted and observed stability. Numerical calculations predict the transverse stiffness of the Bessel beam comparable to that of a focused beam (Figs.~\ref{fig:ForcesBeta20}, S3, S13 and S14 in Ref.~\cite{SM}), our experiments reveal higher stability for the Bessel beam. 
This discrepancy is not accounted for by other considerable transverse forces. For example, the lift force from acoustic streaming is negligible ($\sim10^{-9}\,\text{N}$, see details in Ref.~\cite{SM}), and the stable trapping of non-spherical, non-rotating particles rules out significant symmetry dependent forces (Movies~S3, S4, and S11 in Ref.~\cite{SM}). Further our analysis shows that, unlike a focused beam, the Bessel beam sustains restoring transverse radiation force, well above its axial equilibrium point, indicating a larger levitation-capable region in the axial direction (Fig.~S14 in Ref.~\cite{SM}). Yet the physical mechanism responsible for this enhanced stability remains an open question and may involve factors not captured by our model.

% 日本語コメント：【多段トラップ現象の謎】
Moreover, numerical model predicts existence of a single axial equilibrium point [Fig.~\ref{fig:ForcesBeta20}(e)]. However, we observed particles transitioning between multiple discrete levitation heights in experiments (Fig.~S10 in Ref.~\cite{SM}). The spacing between these heights is approximately half a wavelength ($\frac{\lambda}{2} = 4.26~\text{mm}$). Our pressure field measurements (Fig.~S5 in Ref.~\cite{SM}) confirm that the constructed field retains the characteristics of a traveling wave and matches numerical simulations, thereby ruling out the formation of zero-pressure nodes. Interestingly, a simple model including a weak reflection from the ceiling of only 6 Pa ($<1$\% of the incident beam), corresponding to the PAT being positioned $\sim$1.55 m ($182\lambda$) from the ceiling (Fig.~S15 in Ref.~\cite{SM}), qualitatively reproduces the emergence of the discrete levitation heights (albeit a more complete reflection model is necessary to fully evaluate the effect of reflections on trap stability). This finding demonstrates that accounting for reflections, even when their amplitudes are much smaller than the incident wave, is essential for the future design of single-sided levitation systems. 

% 日本語コメント：【結論と展望】本研究は、片側からの非接触操作を可能にする新たなプラットフォームを確立すると同時に、その過程でモデル化されていない物理現象の豊かな土壌を明らかにしました。これらの複雑なダイナミクスを支配する音響放射力、音響ストリーミング、粒子散乱の相互作用を解明することは、基礎音響学と応用マニピュレーションの両分野における今後の研究の重要な課題となります。
These results establish a novel mode of acoustic levitation, distinct from previous methods based on entrapping particles between high-pressure regions, such as at the nodes of a standing wave or in the central minima of twin and vortex traps. The unique advantages of this high-pressure trapping, including its robustness against obstacles and longer working distance, establish a new platform for versatile non-contact manipulation in an open environment.

\vspace{\baselineskip}
{\itshape Acknowledgments—}This work was supported by Japan Science and Technology Agency (JST) as part of Adopting Sustainable Partnerships for Innovative Research Ecosystem (ASPIRE) Grant Number JPMJAP2330, JST SPRING Grant Number JPMJSP2124, Japan Society for the Promotion of Science (JSPS) KAKENHI Grant Number 25KJ0673, and Engineering and Physical Sciences Research Council (EPSRC) Grant Number EP/Z534171/1. We used OpenAI’s GPT-o3, GPT-5, and Google’s Gemini 2.5 Pro for light language editing.

\vspace{\baselineskip}
{\itshape Data availability—}The data that support the findings of this Letter are openly available \cite{Koroyasu2025Dataset}.
%TC:endignore

\appendix

\nocite{*}

\bibliography{mybib}% Produces the bibliography via BibTeX.

\end{document}

\title{Supplementary Material for ``Mid-Air Single-Sided Acoustic Levitation in High-Pressure Regions of Zero-Order Bessel Beams''}

\author{Yusuke Koroyasu}
\email{koroyu@digitalnature.slis.tsukuba.ac.jp} % 対応著者
\affiliation{Graduate School of Comprehensive Human Science, University of Tsukuba, Tsukuba, Ibaraki 305-8550, Japan}

\author{Christopher Stone}
\affiliation{School of Electrical, Electronic and Mechanical Engineering, University of Bristol, Bristol BS8 1TR, United Kingdom}

\author{Yoichi Ochiai}
\affiliation{Pixie Dust Technologies, Inc., Chuo-ku, Tokyo 104-0028, Japan}
\affiliation{R\&D Center for Digital Nature, University of Tsukuba, Tsukuba, Ibaraki 305-8550, Japan}
\affiliation{Institute of Library, Information and Media Science, University of Tsukuba, Tsukuba, Ibaraki 305-9550, Japan}
\affiliation{Tsukuba Institute for Advanced Research (TIAR), University of Tsukuba, Tsukuba, Ibaraki 305-8577, Japan}

\author{Takayuki Hoshi}
\affiliation{Pixie Dust Technologies, Inc., Chuo-ku, Tokyo 104-0028, Japan}

\author{Bruce W. Drinkwater}
\affiliation{School of Electrical, Electronic and Mechanical Engineering, University of Bristol, Bristol BS8 1TR, United Kingdom}

\author{Tatsuki Fushimi}
\affiliation{R\&D Center for Digital Nature, University of Tsukuba, Tsukuba, Ibaraki 305-8550, Japan}
\affiliation{Institute of Library, Information and Media Science, University of Tsukuba, Tsukuba, Ibaraki 305-9550, Japan}
\affiliation{Tsukuba Institute for Advanced Research (TIAR), University of Tsukuba, Tsukuba, Ibaraki 305-8577, Japan}

%\date{\today}

\maketitle

% 日本語コメント：このセクションでは、本研究で用いた実験装置の基本的な構成について詳述します。
\section{Basic Experimental Setup}
\subsection{Experimental Configuration}
\begin{figure}[b]
    \centering
    \includegraphics[width=0.95\textwidth]{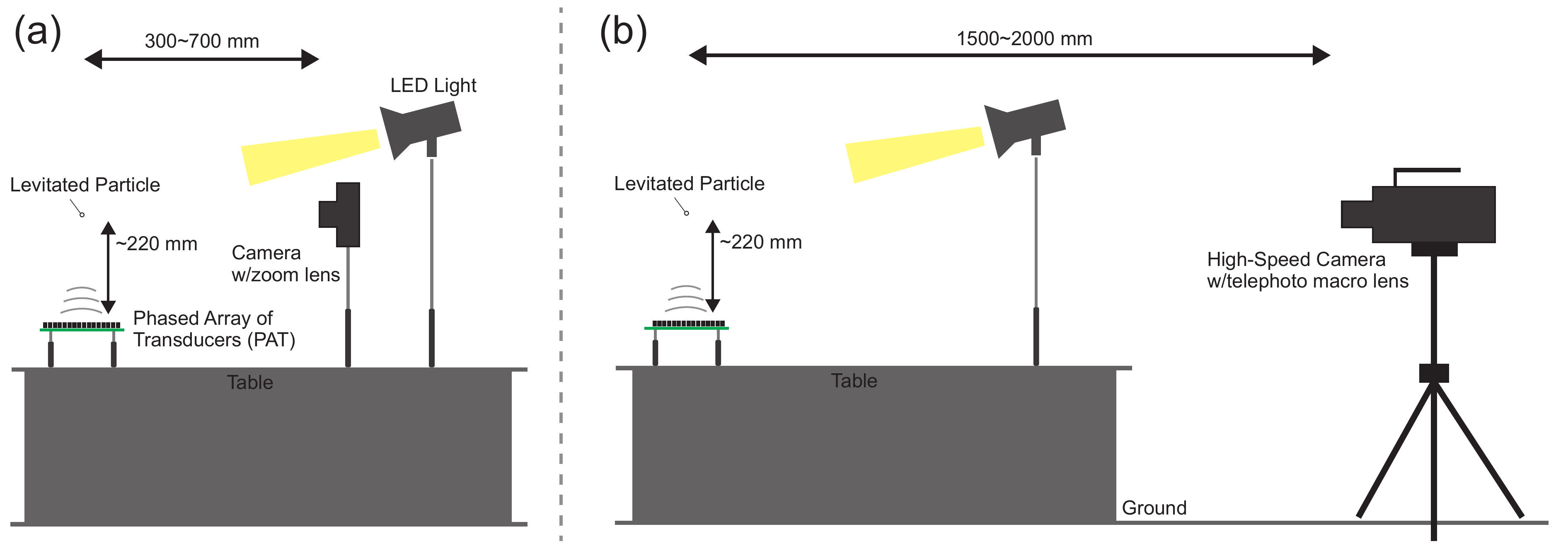}
    \caption{Experimental configurations for side-view imaging: (a) mirrorless camera with a zoom lens; (b) high-speed camera with a telephoto macro lens.}
    \label{fig:experimental_setup}
\end{figure}
% 日本語コメント：サプマテ用—既存原稿のトーンに整合させた装置・撮像の要約（重複排除・図参照）
A \(16\times 16\) phased array of transducers (PAT) generated a zero-order Bessel beam, and a horizontally aligned camera was used to capture the particle position in the vertical plane [Fig.~\ref{fig:experimental_setup}]. The PAT was positioned approximately 1550 mm ($182\lambda$) from its top surface to the ceiling. The array was built following the open-hardware design of Morales \textit{et al.} and the SonicSurface documentation~\cite{morales2021generating,SonicSurface_Instructables}, using Murata MA40S4S transducers (\(40~\mathrm{kHz}\)). The PAT was driven by MIC4127 MOSFET drivers controlled by an Altera CoreEP4CE6 FPGA; serial communication used a USB FTDI link operating at \(230{,}400\) bps. The array was powered by a programmable DC supply (Kikusui PWR801L), and to ensure consistent conditions the PAT was powered for approximately one hour before the experiments to reach thermal equilibrium.

Unless otherwise stated, all experiments were conducted with a primary expanded polystyrene (EPS) sphere of radius $a=0.75\,\text{mm}$ and density $\rho_p=40.4\,\text{kg\,m}^{-3}$. These properties were derived from dimension and mass measurements using a digital microscope and an ultramicrobalance (Sartorius MSA2.7S-000-DM), respectively.

% 日本語コメント：このサブセクションでは、本研究で用いた全ての音響トラップを生成するための位相計算、およびビームを電子的に走査する手法について具体的に記述します。
\subsection{Phase Calculation}
This section details the phase calculations for acoustic holograms used in this study. For a standard focused beam, the phase $\phi_{i, \text{focus}}$ for each transducer at position $\boldsymbol{x}_i=(x_i, y_i, z_i)$ is set to focus at $\boldsymbol{x}_f=(x_f, y_f, z_f)$:
\begin{equation}
    \phi_{i, \text{focus}} = -k (|\boldsymbol{x}_i - \boldsymbol{x}_f| - |\boldsymbol{x}_f|),
    \label{eq:sm_focus_phase}
\end{equation}
where $k = \frac{2\pi f}{c_0}$ is the wavenumber at a frequency of $f=40\,\text{kHz}$ and speed of sound in air $c_0 = 341\,\text{m\,s}^{-1}$.
The twin trap, used for the working range comparison (Fig.~2(g) in the main text), was created by modifying this focal phase. A phase shift of $\pi$ was added to the transducers located in the half-space where the $x$-coordinate $x_i \leq 0$~\cite{marzo2015holographic}. The resulting phase profile, $\phi_{i, \text{twin}}$, is given by:
\begin{equation}
\phi_{i, \text{twin}} =
    \begin{cases}
        \phi_{i, \text{focus}} + \pi, & x_i \leq 0 \\
        \phi_{i, \text{focus}}, & \text{otherwise}.
    \end{cases}
    \label{eq:sm_twin_phase}
\end{equation}

\begin{figure}[t]
    \centering
    \includegraphics[width=0.6\textwidth]{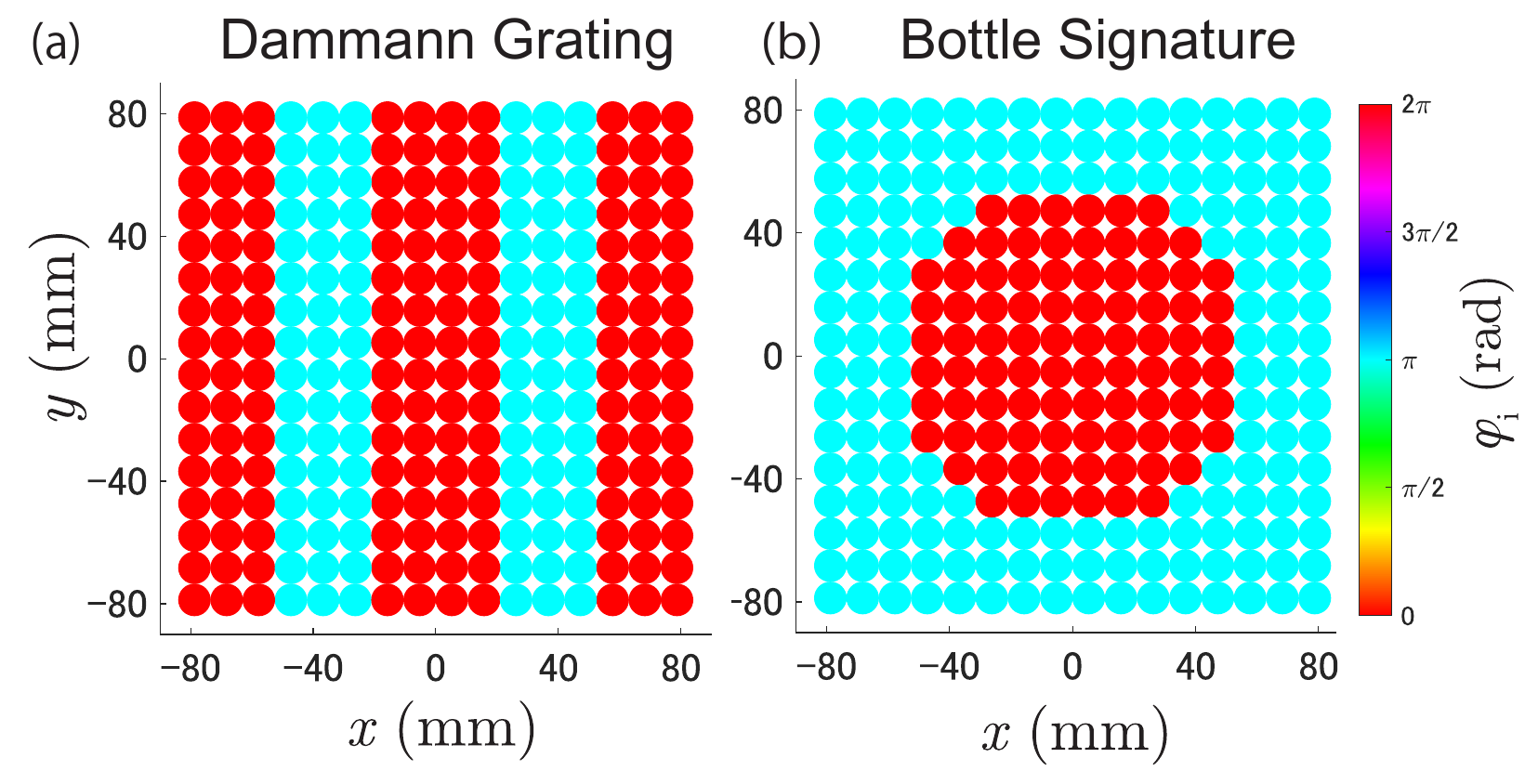}
    \caption{Phase signatures applied to the primary Bessel beam hologram. The red color indicates no shift, while the cyan indicates a phase shift of $\pi$. (a) Dammann grating pattern. (b) Bottle signature pattern.}
    \label{fig:binary_modulations}
\end{figure}

The phase hologram for the primary levitation field, the zero-order Bessel beam, is given by~\cite{hasegawa2017electronically, norasikin2019sonicspray}:
\begin{equation}
\phi_{i, \text{Bessel}} = -k( \sin\beta\sqrt{x_i^2+y_i^2} - \cos\beta z_i).
\label{eq:sm_bessel_phase}
\end{equation}
To tilt this beam in the $x$-$z$ plane, the transducer positions $\boldsymbol{x}_i$ were first virtually rotated about the axis $\boldsymbol{v}_{\text{rot}} = (0, 1, 0)$ by an angle $\theta_{\text{tilt}}$. The rotated position of each transducer, $\boldsymbol{x}_{i,\mathrm{rot}}$, is calculated using Rodrigues' rotation formula:
\begin{equation}
    \boldsymbol{x}_{i,\mathrm{rot}} = \cos \theta_{\text{tilt}}\, \boldsymbol{x}_i + \sin \theta_{\text{tilt}}\, (\boldsymbol{v}_{\text{rot}} \times \boldsymbol{x}_i) + (1 - \cos \theta_{\text{tilt}})(\boldsymbol{v}_{\text{rot}} \cdot \boldsymbol{x}_i)\, \boldsymbol{v}_{\text{rot}}.
    \label{rodrigues_rotation_modified}
\end{equation}
The final phase for the tilted beam is then calculated by substituting the components of the rotated position vector $\boldsymbol{x}_{i,\mathrm{rot}}$ into $\boldsymbol{x}_i$ of Eq.~\eqref{eq:sm_bessel_phase}~\cite{hasegawa2017electronically}. 

% 日本語コメント：複数粒子操作のための追加ホログラムについて、ベッセルビームの基本ホログラムに位相変調を加える形で記述する。具体的に、Dammannグレーティングとボトルシグネチャの最終的な位相計算式をそれぞれ提示し、変調領域を規定する遷移点を定義する。最後に、これらの遷移点の調整による粒子間隔の操作可能性について言及する。
Additional holograms for multi-particle manipulation were created by modifying the Bessel beam hologram, $\phi_{i, \text{Bessel}}$. The phase for the Dammann grating, $\phi_{i, \text{Dammann}}$, used for parallel alignment [Fig.~\ref{fig:binary_modulations}(a)], is calculated as follows:
\begin{equation}
\phi_{i, \text{Dammann}} = 
    \begin{cases}
        \phi_{i, \text{Bessel}} + \pi, & X_1 < |x_i| < X_2 \\
        \phi_{i, \text{Bessel}}, & \text{otherwise}
    \end{cases}
    \label{eq:sm_dammann_phase}
\end{equation}
where the transition points used in this study are $X_1=0.1A$ and $X_2=0.3A$, with the array size $A=160\,\text{mm}$. The phase for the bottle signature, $\phi_{i, \text{bottle}}$, used for vertical alignment [Fig.~\ref{fig:binary_modulations}(b)], is calculated as:
\begin{equation}
\phi_{i, \text{bottle}} = 
    \begin{cases}
        \phi_{i, \text{Bessel}} + \pi, & \sqrt{x_i^2+y_i^2} > R_B \\
        \phi_{i, \text{Bessel}}, & \text{otherwise}
    \end{cases}
    \label{eq:sm_bottle_phase}
\end{equation}
where the radial transition point is $R_B = 0.35A$. By adjusting these transition points, the spacing between the pressure peaks can be controlled.

% 日本語コメント：このサブセクションでは、本研究の実験結果を記録するために使用した全ての撮影機材、レンズ、および記録設定を網羅的に記述します。各機材が、本文および補足資料中のどの図や動画に対応するかを明確に紐付けることで、研究の再現性を担保します。
\subsection{Recording Devices}
The experimental results were recorded using several camera systems. The majority of the high-speed recordings, including Figs.~2(a), 2(d), 3, and 4(a) in the main text, Figs.~S12, and Movies~S5 (side view), S6, S7, S8, S9, S10, and S11, were captured using a high-speed camera (FASTCAM Nova R2, Photron) with a telephoto macro lens (Tamron SP AF180 mm F/3.5 Di LD [IF] MACRO 1:1). These were recorded at 125 frames per second (fps) with a resolution of \(2048 \times 2048\) pixels, with the exception of Movie S11, which was captured at 2000 fps with a resolution of \(1024 \times 1024\) pixels. Additional recordings were made with a mirrorless camera (ILCE-9, Sony), used with two different lenses. With a Vario-Tessar T\textsuperscript{*} FE 16–35 mm F4 ZA OSS lens, it was used for Fig.~S10 and Movies S1 and S2, while a FE 24-105 mm F4 G OSS lens was used for Fig.~4(b) in the main text and Movies S3, S4, and S5 (top view). Both configurations recorded at 240 fps with a resolution of \(1920 \times 1080\) pixels. For the working range comparison shown in Fig.~2(g) of the main text, a USB camera (Infinicam, Photron) with a telecentric lens (KCM-0914MP5, Tokina) was employed at a resolution of \(1246 \times 1008\) pixels. The lenses used to detect the position of the particles were calibrated using a checkerboard pattern.

% 日本語コメント：このセクションでは、本文の議論の前提となる予備実験について記述します。
\section{Preliminary Experiment: Stability of a Focused Beam Trap}
\begin{figure}[b]
    \centering
    \includegraphics[width=0.79\textwidth]{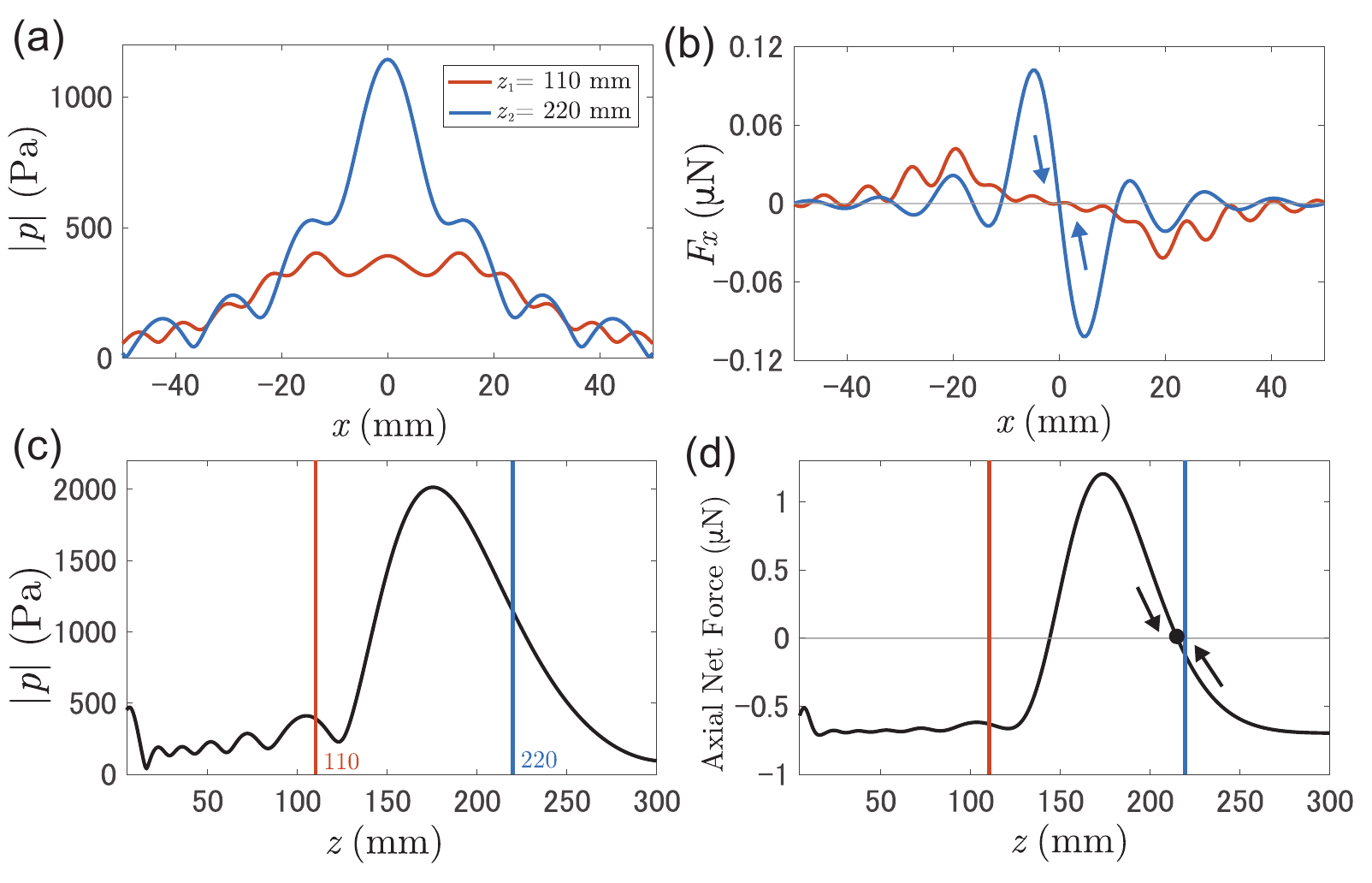}
    \caption{Calculated acoustic pressure field of a focused beam from the PAT used in preliminary experiments ($z_f = 180$ mm, $V_\text{in}=8\,\text{V}$) and the radiation force on an EPS sphere (radius $a=0.75\,\text{mm}$). (a) Transverse and (c) axial pressure profiles. (b) Transverse force $F_x$. For the axial plots (c) and (d), solid vertical lines indicate the two evaluation planes, \(z_1=110\,\text{mm}\) (red) and \(z_2=220\,\text{mm}\) (blue, near the experimental levitation height). (d) Net axial force. The black line shows the balance between radiation force and gravity.}
    \label{fig:focus_180mm}
\end{figure}

A preliminary experiment was conducted to experimentally validate the difficulty of achieving stable levitation at a high-pressure focal point. These focused beam trials were performed alternated with the Bessel beam experiments [Fig.~\ref{fig:SI_t_vs_z}] to ensure identical ambient conditions for a direct comparison. An EPS sphere (radius \( a = 0.75 \) mm) was placed on an acoustically transparent mesh, and the PAT was configured to generate a single focused beam at a height of \( z_f = 180 \) mm ($21.1\lambda$). This focal length was chosen to create an experimental levitation height of approximately 220 mm, similar to that achieved with the \( \beta = 20^\circ \) Bessel beam with same input voltage ($V_{in}=8$ V). Numerical calculations predict that this beam forms a stable three-dimentional trap with restoring forces in both the transverse and axial directions, creating an equilibrium point at approximately $z=215$ mm ($25.2\lambda$) [Fig.~\ref{fig:focus_180mm}]. Levitation was initiated by manually vibrating the mesh. Once the particle was suspended, the mesh was carefully withdrawn to observe the stability of the trap.

\begin{figure}[t]
    \centering
    \includegraphics[width=0.79\textwidth]{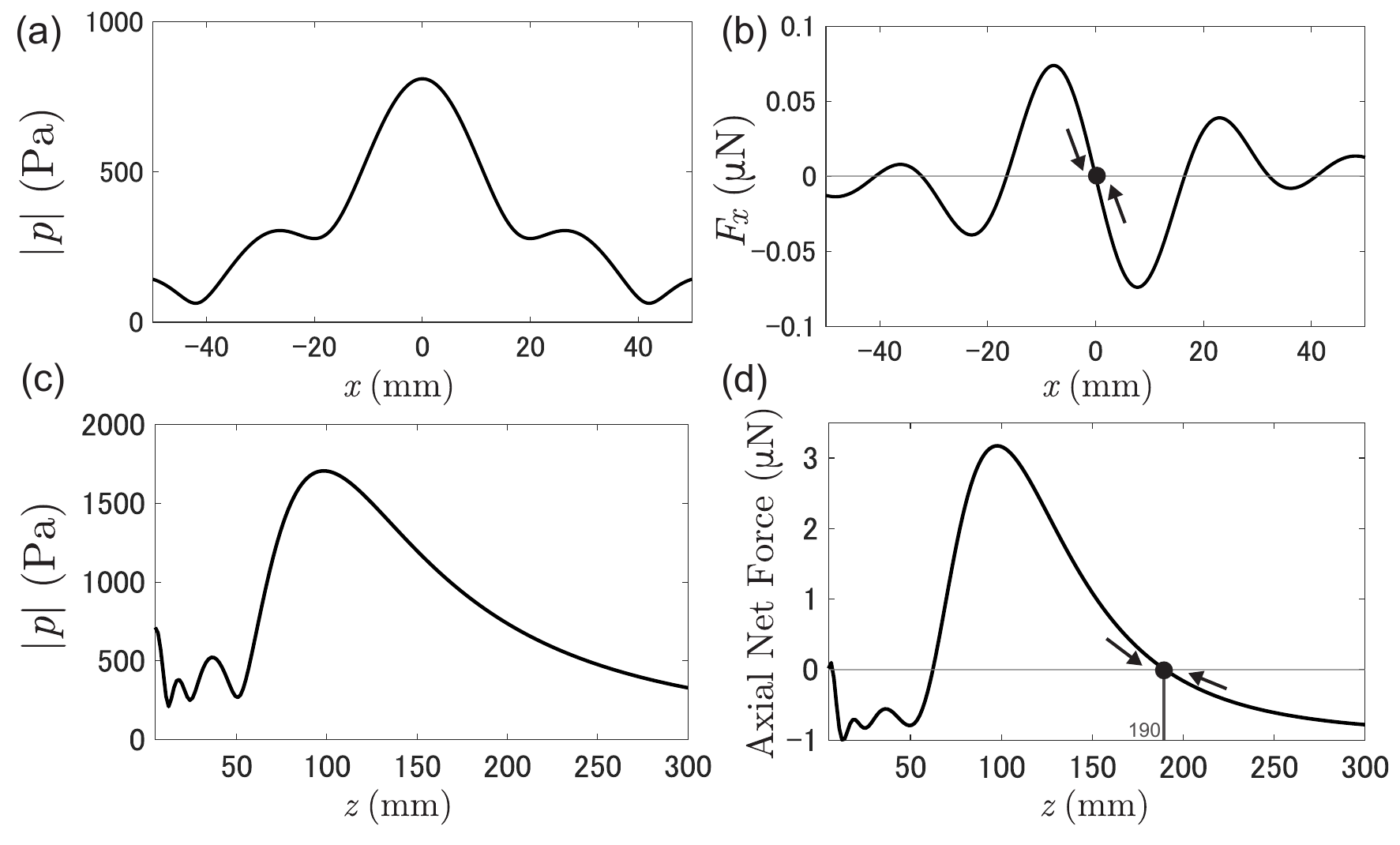}
    \caption{Verification of the three-dimensional restoring radiation force from the focused beam in Ref.~\cite{marzo2015holographic}. The sound pressure and radiation force profiles are shown for an expanded polystyrene sphere ($a=0.92\,\text{mm}$ ($\frac{a}{\lambda}=0.11$), $\rho_p=29.36\,\text{kg\,m}^{-3}$) with beam parameters ($z_f = 117$ mm, $V_\text{in}=15\,\text{V}$) estimated from Ref.~\cite{marzo2015holographic}. (a) Transverse pressure profile at $z=190$ mm. (b) Transverse force profile $F_x$ at $z=190$ mm, which acts as a restoring force toward the beam axis. (c) Axial pressure profile. (d) Net axial force (radiation force and gravity), demonstrating a stable equilibrium point where the force is restorative.}
    \label{fig:SI_asier_focus}
\end{figure}

In 15 independent trials on focused beams, the particle was consistently attracted toward the beam axis but failed to remain stably levitated after the mesh was removed. In the most successful trial, levitation was maintained for approximately 3 seconds before the particle was ejected from the trap (see Movie S1). This experimental finding replicates the challenges described by Marzo \textit{et al.}~\cite{marzo2015holographic}. However, our numerical analysis of their configuration shows that the focused beam generates a three-dimensional restoring force [Fig.~\ref{fig:SI_asier_focus}]. For this calculation, we modeled their setup using an $8\cross 8$ PAT. Because the focal position was not specified in Ref.~\cite{marzo2015holographic}, we selected $z_f = 117$ mm ($13.7\lambda$) to create an numerical equilibrium point at approximately 190 mm ($22.3\lambda)$, consistent with reported observations. The analysis confirms a restorative transverse force [Fig.~\ref{fig:SI_asier_focus}(b)] and a stable axial equilibrium point [Fig.~\ref{fig:SI_asier_focus}(d)]. The presence of such a transverse restoring force in a focused beam in mid-air, particularly when the focal point is located far from the PAT (e.g., $z_f>100$ mm), is consistent with Ref.~\cite{koroyasu2023microfluidic}. The difficulty of achieving stable levitation in practice, despite the existence of this theoretically stable trap, motivates the investigation of the Bessel beam presented in the main text.

\newpage

\section{Verification of the generated Bessel beam}
\begin{figure}[h]
    \centering
    \includegraphics[width=0.95\textwidth]{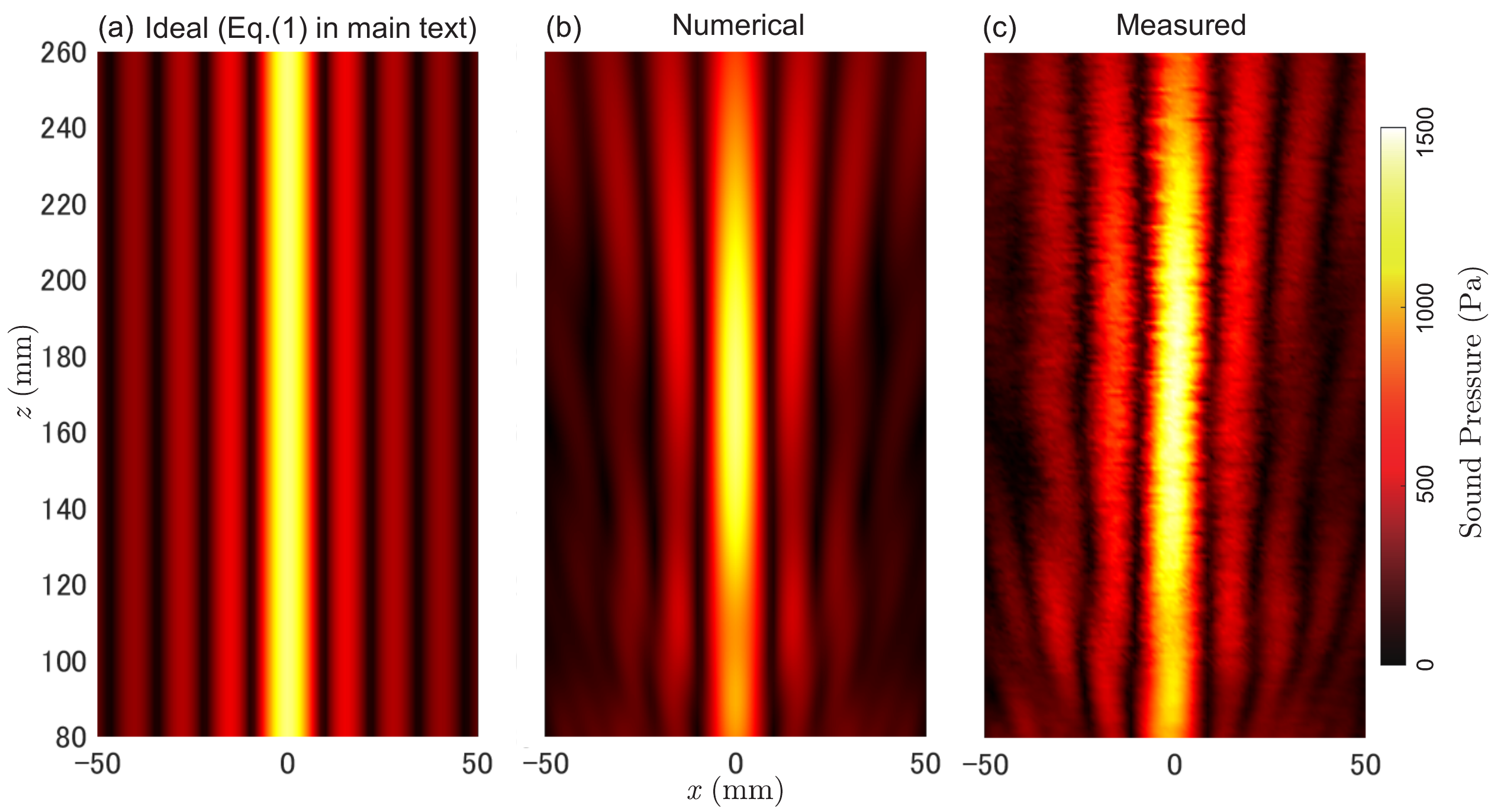}
    \caption{Comparison of zero-order Bessel beam acoustic pressure amplitude in the $xz$-plane for a cone angle of $\beta = 20^\circ$ and an input voltage of $V_\text{in} = 8\,\text{V}$. (a) Ideal pressure field calculated using Eq.~(1) of the main text, assuming an infinite aperture. The pressure amplitude is normalized to the maximum amplitude of the numerical map. (b) Numerically simulated pressure field generated from the finite aperture PAT. (c) Experimentally measured pressure field generated using a PAT. All panels share the same color scale. }
    \label{fig:SI_mic}
\end{figure}

% 日本語コメント：このセクションでは、実験的に生成したベッセルビームが数値計算と一致することを検証します。まず音場の測定手法を具体的に記述し、次に図を用いて測定結果と計算結果を比較・提示することで、実験の妥当性を示します。
To measure the pressure distribution, a calibrated Brüel \& Kjær (B\&K) 4138-A-015 1/8-inch microphone was mounted on a three-axis linear stage (Ossila). The microphone was then scanned across the $xz$-plane with a spatial resolution of 1 mm. Figure~\ref{fig:SI_mic} shows a comparison for a zero-order Bessel beam generated with a cone angle of $\beta = 20^\circ$ and an input voltage of $V_\text{in} = 8\,\text{V}$. The figure compares the experimentally measured pressure amplitude [Fig.~\ref{fig:SI_mic}(c)] with the ideal field calculated using Eq.~(1) of the main text assuming an infinite aperture [Fig.~\ref{fig:SI_mic}(a)], and a numerical simulation from our finite aperture PAT [Fig.~\ref{fig:SI_mic}(b)]. The measured field profile exhibits the characteristic high-pressure axial core and concentric side rings and shows good agreement with the simulated field from the finite PAT. Importantly, the profile does not have the zero-pressure nodes. These observations confirm the experimental generation of the zero-order Bessel beam.

\newpage
% =================================================================================
% 日本語コメント：このセクションでは、本研究で用いた音響放射力の数値計算に関する手法を詳述します。
\section{Numerical Calculations}

% 日本語コメント：このサブセクションでは、音響放射力の計算の基礎となる、任意の点での音圧の計算法について記述します。
\subsection{Total Pressure Calculations}
The acoustic pressure field was calculated based on Huygens' principle as follows:
\begin{equation}
p(\boldsymbol{x}) = \sum_{i=1}^{T} \frac{P_0}{|\boldsymbol{x} - \boldsymbol{x}_i|}\, D(\eta_i)\, e^{j\left( k|\boldsymbol{x} - \boldsymbol{x}_i| + \phi_i \right)},
\label{eq:sm_pressure}
\end{equation}
where \(p(\boldsymbol{x})\) is the complex acoustic pressure at position \(\boldsymbol{x}\), \(T=256\) is the total number of transducers, \(\boldsymbol{x}_i\) is the position of the \(i\)-th transducer, \(\phi_i\) is its phase, and \(j\) is the imaginary unit. The source amplitude \(P_0\) is related to the input voltage \(V_{\text{in}}\) by \(P_0 = 0.221 V_{\text{in}}\) \(\text{Pa}\,\text{m}\). The term \(D(\eta_i) = \frac{2 J_1(k a_t \sin \eta_i)}{k a_t \sin \eta_i}\) denotes the far-field directivity function of a piston source, with \(J_1\) being the first-order Bessel function of the first kind, \(a_t\) the radius of the transducer, and \(\eta_i\) the angle between the normal of the \(i\)-th transducer and the vector \(\boldsymbol{x} - \boldsymbol{x}_i\). 

% 日本語コメント：このサブセクションでは、本研究で用いられた音響放射力の計算手法について詳述します。
\subsection{Acoustic Radiation Force Calculation}
%【段落の役割】レイリー領域における半径aの圧縮性球体への音響放射力を、Bareschらの理論式を用いて定義する。式の後にwhereを用いて各物理量を定義し、式の各項の物理的意味と修正点を簡潔に説明することで、本研究で用いる理論的枠組みを示す。
The acoustic radiation force on a compressible sphere of radius $a$ in the Rayleigh regime ($a \ll \lambda$) is given by the formulation of Baresch et al.~\cite{baresch2016observation}:
\begin{equation}
\label{eq:BareschForce}
\begin{aligned}
\boldsymbol{F}
&= -\frac{\mathrm{Re}(\alpha_m)\,\kappa_0}{4}\,\nabla |p|^2
\;+\; \frac{\mathrm{Re}(\alpha_d)\,\rho_0}{4}\,\nabla |\boldsymbol{v}|^2 \\
&\quad -\frac{1}{2}\Big\{
\big[\tfrac{k}{c_0}\mathrm{Im}(\alpha_m)-\tfrac{k^{4}}{6\pi c_0}\mathrm{Re}(\alpha_m)\mathrm{Re}(\alpha_d)\big]\,
\mathrm{Re}\!\big(p\,\boldsymbol{v}^{*}\big)
\;+\; \rho_0\,\mathrm{Im}(\alpha_d)\,\mathrm{Im}\!\big[(\boldsymbol{v}\!\cdot\!\nabla)\,\boldsymbol{v}^{*}\big]
\Big\},
\end{aligned}
\end{equation}
where $p$ and $\boldsymbol{v}= \frac{1}{j\,\omega\,\rho_0}\,\nabla p$ are the complex acoustic pressure and particle velocity, respectively. The surrounding medium is air with density $\rho_0 = 1.21\,\text{kg\,m}^{-3}$. The asterisk (\(*\)) denotes the complex conjugate, and $\mathrm{Re}(\cdot)$ and $\mathrm{Im}(\cdot)$ represent the real and imaginary parts. The terms $\alpha_m$ and $\alpha_d$ are the acoustic strength parameters,
\begin{equation}
\alpha_m=\frac{\alpha_m^{0}}{1+j\,\frac{k^{3}}{4\pi}\,\alpha_m^{0}},
\qquad
\alpha_d=\frac{\alpha_d^{0}}{1-j\,\frac{k^{3}}{12\pi}\,\alpha_d^{0}}.
\end{equation}
The parameters $\alpha_m^0$ and $\alpha_d^0$ are the sphere's monopolar and dipolar modes given by:
\begin{equation}
\alpha_m^{0}=V_p\!\left(\,1-\frac{\kappa_p}{\kappa_0}\,\right),
\qquad
\alpha_d^{0}=3V_p\!\left(\frac{\rho_p-\rho_0}{2\rho_p+\rho_0}\right),
\end{equation}
where $V_p = \frac{4}{3}\pi a^3$ is the particle volume. The particle is modeled as a compressible sphere with density $\rho_p=40.4\,\text{kg\,m}^{-3}$ and longitudinal propagation speed $c_p=900~\mathrm{m\,s^{-1}}$. The compressibilities of the medium and the particle are $\kappa_0 = \frac{1}{\rho_0 c_0^2}$ and $\kappa_p = \frac{1}{\rho_p c_p^2}$, respectively. The first two terms on the right-hand side of Eq.~\eqref{eq:BareschForce} represent the gradient force, while the remaining terms form the scattering force. The term in the scattering force containing $\mathrm{Re}(\alpha_m)\mathrm{Re}(\alpha_d)$ incorporates a correction for a missing factor of 2 in the original expression of Ref.~\cite{baresch2016observation}, as pointed out in Ref.~\cite{gong2022single}.

%【段落の役割】Bareschの式が、散乱力が無視できる条件下で、著名なGor'kovポテンシャルに単純化されることを示す。
When scattering force is negligible, Eq.~\eqref{eq:BareschForce} simplifies to Gor'kov potential~\cite{gorkov1961dokl,bruus2012acoustofluidics}. The acoustic radiation force is then derived from the time-averaged potential:
\begin{equation}
U = V_p \left( \frac{1}{2} \kappa_0 f_1 \langle p^2 \rangle - \frac{3}{4} \rho_0 f_2 \langle \boldsymbol{v}^2 \rangle \right),
\label{eq:gorkov_potential}
\end{equation}
where $\langle \cdot \rangle$ denotes a time average. The monopole coefficient $f_1$ and dipole coefficient $f_2$ are defined as $f_1 = 1 - \frac{\kappa_p}{\kappa_0}$ and $f_2 = \frac{2(\rho_p - \rho_0)}{2\rho_p + \rho_0}$. The force is then given by $\boldsymbol{F} = -\nabla U$.

%【段落の役割】本研究の定量的計算で主に使用した、AnderssonとAhrensによる音響放射力の計算手法を導入する。その手法がPATによる音場計算に適しており、かつレイリー領域に限定されないという利点を述べ、その基本原理と適用範囲を簡潔に説明することで計算の妥当性を示す。
Unless otherwise stated, we computed the acoustic radiation force using the Andersson and Ahrens method, which is well-suited to PAT-generated fields and remains valid beyond the Rayleigh regime~\cite{Andersson2019, Andersson2022, zehnter2021acoustic}. In this framework, both the incident field, modeled as a superposition of $T$ piston sources with prescribed phases, and the field scattered by a compressible sphere are expanded in spherical harmonics. The force components are then given by:
\begin{align}
F_x &= \frac{1}{8\rho_0 c_0^2 k^2}\, \operatorname{Re} \left\{ \sum_{n=0}^{\infty} \sum_{m=-n}^{n} \Psi_n A_{nm} \left( S_{nm}\, S^{*}_{n+1, m+1} - S_{n, -m}\, S^{*}_{n+1, -m-1} \right) \right\}, \label{eq:Fx} \\
F_y &= \frac{1}{8\rho_0 c_0^2 k^2}\, \operatorname{Im} \left\{ \sum_{n=0}^{\infty} \sum_{m=-n}^{n} \Psi_n A_{nm} \left( S_{nm}\, S^{*}_{n+1, m+1} + S_{n, -m}\, S^{*}_{n+1, -m-1} \right) \right\}, \label{eq:Fy} \\
F_z &= \frac{1}{8\rho_0 c_0^2 k^2}\, \operatorname{Re} \left\{ \sum_{n=0}^{\infty} \sum_{m=-n}^{n} \Psi_n B_{nm}\, S_{nm}\, S^{*}_{n+1, m} \right\}. \label{eq:Fz}
\end{align}
where the coefficients $\Psi_n$, $A_{nm}$, and $B_{nm}$ are defined as:
\begin{align}
\Psi_n &= 2j \left( c_n + c^{*}_{n+1} + 2 c_n c^{*}_{n+1} \right), \\
A_{n m} &= \sqrt{\frac{(n + m + 1)(n + m + 2)}{(2n + 1)(2n + 3)}}, \\
B_{n m} &= -2 \sqrt{\frac{(n + m + 1)(n - m + 1)}{(2n + 1)(2n + 3)}}.
\end{align}
The total beam shape coefficients for the incident field, $S_{nm}$, are calculated by summing the contributions from each of the transducers~\cite{zehnter2021acoustic}:
\begin{equation}
S_{n m} = \sum_{i=1}^{T} 4\pi P_0\, e^{j \phi_i}\, D(\eta_i)\, j k h^{(1)}_n(k r_i)\, Y^{*}_{n m}(\theta_i, \varphi_i),
\end{equation}
where \(Y_{nm}(\theta,\varphi)\) are the complex spherical harmonics, \(h_n^{(1)}\) denotes the spherical Hankel function of the first kind, and \((r_i,\theta_i,\varphi_i)\) are the spherical coordinates of the \(i\)-th transducer measured from the center of the sphere. The scattering coefficient for a compressible sphere, $c_n$, is given by:
\begin{equation}
c_n = -\frac{j_n(k a)\, j^{\prime}_{n}\left(k_p a\right) - Z\, j^{\prime}_{n}\left(k a\right)\, j_n\left(k_p a\right)}{h^{(1)}_n(k a)\, j^{\prime}_{n}\left(k_p a\right) - Z\, h^{(1)\prime}_n(k a)\, j_n\left(k_p a\right)},
\end{equation}
where $Z = \frac{\rho_p c_p}{\rho_0 c_0}$ is the relative impedance, the prime $\prime$ indicates the derivative of a function with respect to its argument, and $j_n$ denotes the spherical Bessel functions of the first kind. In the numerical implementation, the infinite summations in Eqs.~\eqref{eq:Fx}-\eqref{eq:Fz} are performed iteratively starting from $n=0$ and are truncated once the relative change in the total force between successive orders falls below the predefined tolerance of $1 \times 10^{-5}$, ensuring numerical convergence.

\begin{figure}[t]
    \centering
    \includegraphics[width=0.65\textwidth]{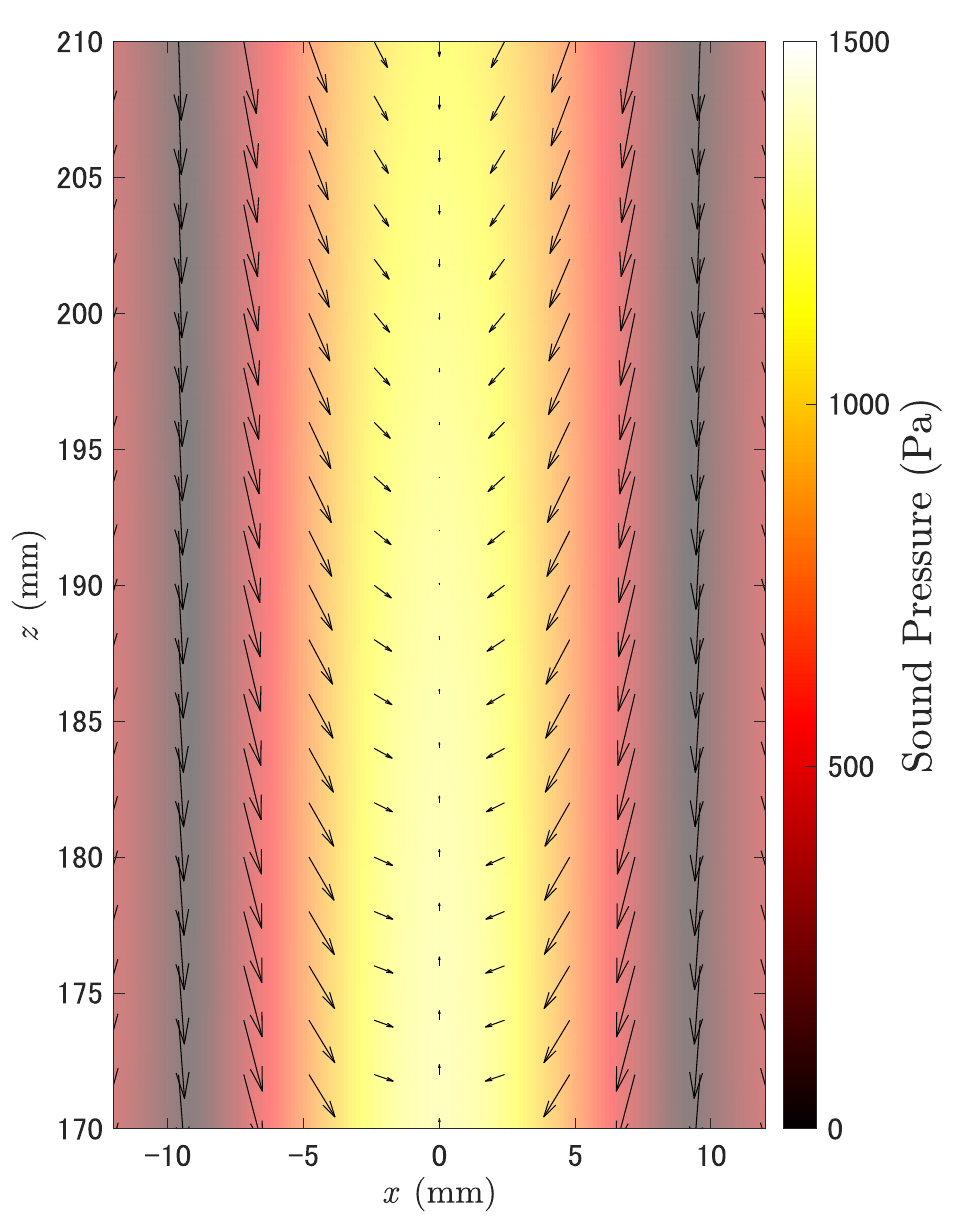}
    \caption{Net force field (acoustic radiation force and gravity) on an EPS sphere ($a=0.75\,\text{mm}$) in the $xz$-plane for a zero-order Bessel beam ($\beta = 20^\circ$, $V_\text{in}=8\,\text{V}$). The background shows pressure amplitude, and arrows represent the net force vectors.}
    \label{fig:SI_quiver}
\end{figure}

% 日本語コメント：この段落では、音響放射力場を可視化し、トラップメカニズムを定性的に説明する。
Figure~\ref{fig:SI_quiver} shows the net force field (acoustic radiation force and gravity) in the $xz$-plane, overlaid on the pressure amplitude. The acoustic radiation force is calculated using Andersson and Ahrens method [Eqs.~\eqref{eq:Fx}-\eqref{eq:Fz}] for an EPS sphere ($a=0.75\,\text{mm}$) in a zero-order Bessel beam ($\beta = 20^\circ$, $V_\text{in}=8\,\text{V}$). The vector field indicates a stable three-dimensional trap: lateral restoring forces direct particles towards the high-pressure beam axis, and the axial component balances gravity at an equilibrium position. This visualization is consistent with the force profiles presented in the main text [Figs.~1(c) and 1(e)].

\newpage

% 日本語コメント：このサブセクションでは、本研究で用いた音響放射力の数値計算コードの妥当性を検証するプロセスについて記述します。
\subsection{Validation of Acoustic Radiation Force Calculation}
\begin{figure}[t]
    \centering
    \includegraphics[width=0.8\textwidth]{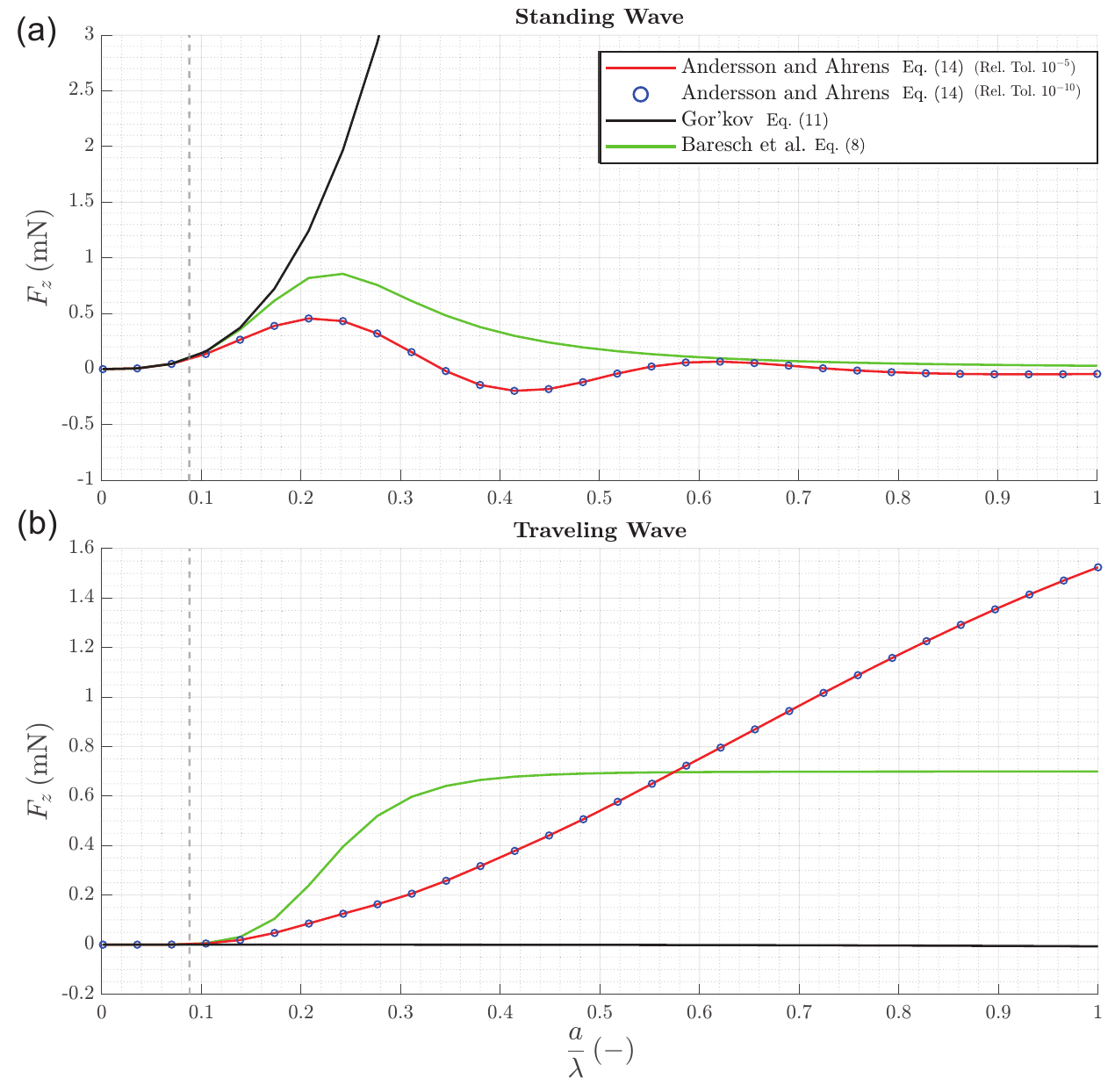}
    % 日本語コメント：図S11のキャプション—焦点座標を (x_f,y_f,z_f)=(0,0,121) mm に明示し，用語・記法を本文と統一
    \caption{Comparison of the axial acoustic radiation force as a function of \(\frac{a}{\lambda}\). (a) Standing wave: evaluated at \(z=120\,\mathrm{mm}\) for two opposing PATs with the focus set at \((x_f,y_f,z_f)=(0,0,121)\,\mathrm{mm}\). (b) Traveling wave: single PAT, evaluated at \(z=z_f=180\,\mathrm{mm}\). Plots show Andersson and Ahrens [Eqs.~\eqref{eq:Fx}–\eqref{eq:Fz}], Gor'kov potential [Eq.~\eqref{eq:gorkov_potential}], and Baresch model [Eq.~\eqref{eq:BareschForce}]. Results with summation tolerances \(1\times10^{-5}\) (red solid line) and \(1\times10^{-10}\) (blue circles) overlap. The gray dashed line marks \(\frac{a}{\lambda}\sim 0.088\).}
        \label{fig:SI_Validation_Tol}
\end{figure}

% 日本語コメント：AA法実装の検証—standing wave / traveling wave の2条件で参照理論を切り替える
To validate our implementation of the Andersson and Ahrens formulation [Eqs.~\eqref{eq:Fx}-\eqref{eq:Fz}], we compared model predictions in two distinct acoustic configurations. First, a standing wave field was formed by two opposing \(16\times16\) PATs, positioned at $z_i=0$ and $z_i=240$ mm, with a \(\pi\) phase inversion applied to top array; this serves as a benchmark where the Gor’kov potential [Eq.~\eqref{eq:gorkov_potential}] is accurate. Second, a traveling wave focus generated by a single PAT ($z_i=0$ mm), with the focus set to \(z_f=180\,\mathrm{mm}\), replicated the experimental conditions; in this regime scattering contributions are non-negligible and the Gor’kov potential becomes inadequate~\cite{silva2014acoustic, marzo2018acoustic}. Accordingly, we used the Baresch formulation [Eq.~\eqref{eq:BareschForce}] for comparison in the traveling wave case.

% 日本語コメント：図S11（Fig.~\ref{fig:SI_Validation_Tol}）に基づく粒径依存の比較と数値収束性の記述（\bigl/\bigr の文法エラーを修正）
Figure~\ref{fig:SI_Validation_Tol} compares the axial acoustic radiation force as a function of particle radius. In the standing wave field [Fig.~\ref{fig:SI_Validation_Tol}(a)], the Andersson and Ahrens formulation, the Gor'kov potential [Eq.~\eqref{eq:gorkov_potential}], and the Baresch model [Eq.~\eqref{eq:BareschForce}] agree for \(\frac{a}{\lambda}\lesssim 0.1\). For example, for \(\frac{a}{\lambda}=0.088\), the predicted forces are \(0.087~\mathrm{mN}\) (Andersson and Ahrens), \(0.094~\mathrm{mN}\) (Gor'kov), and \(0.094~\mathrm{mN}\) (Baresch). Beyond this range the Gor'kov potential diverges; although the Baresch model remains closer in the Rayleigh limit, it also degrades for larger radii due to its underlying assumptions. In the traveling wave case [Fig.~\ref{fig:SI_Validation_Tol}(b)], the Gor'kov potential is inadequate across the range, yielding a negative value \((-4.8\times10^{-6}~\mathrm{mN})\) even at \(\frac{a}{\lambda}=0.088\). The Baresch model provides a reasonable estimate in the Rayleigh regime (\(0.0021~\mathrm{mN}\) versus \(0.0019~\mathrm{mN}\) from the Andersson and Ahrens formulation) but becomes unreliable beyond \(\frac{a}{\lambda}\sim 0.1\), exhibiting a maximum near \(\frac{a}{\lambda}\sim 0.4\) that is inconsistent with the expected monotonic increase of force with particle radius. Throughout, a summation tolerance of \(1\times10^{-5}\) produced results indistinguishable from \(1\times10^{-10}\), confirming numerical convergence.

\begin{figure}[H]
    \centering
    \includegraphics[width=0.95\textwidth]{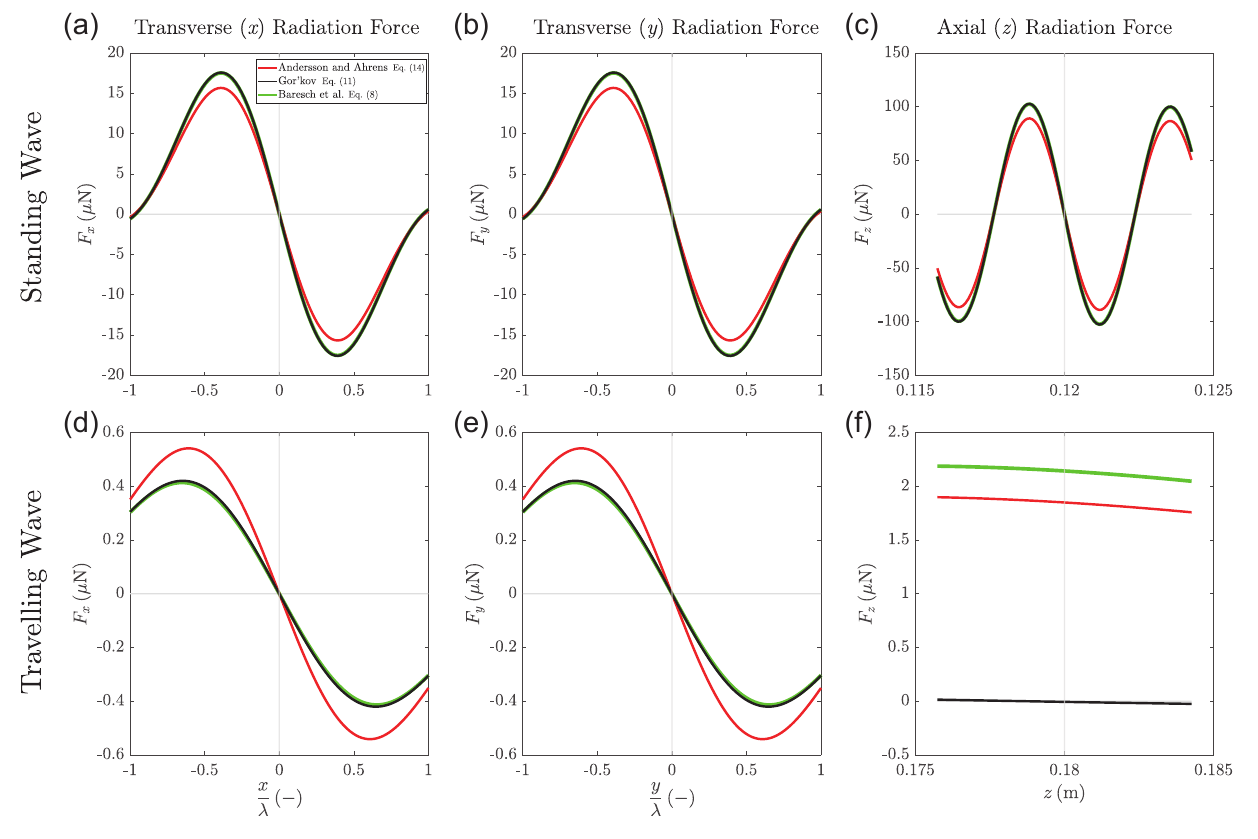}
    \caption{
Spatial distribution of the acoustic radiation force on a particle of radius \(a=0.75\) mm ($\frac{a}{\lambda}= 0.088$), comparing the Andersson and Ahrens method [Eqs.~\eqref{eq:Fx}–\eqref{eq:Fz}], the Gor'kov potential [Eq.~\eqref{eq:gorkov_potential}], and the Baresch formulation [Eq.~\eqref{eq:BareschForce}]. The columns show the force components along the \(x\), \(y\), and \(z\) axes, calculated through the respective focal point for each configuration. (a)–(c) Results for a standing wave generated by two opposing PATs with the focus set at \((0,0,120)\,\mathrm{mm}\). (d)–(f) Results for a traveling wave from a single PAT with the focus set at \((0,0,180)\,\mathrm{mm}\).
}
    \label{fig:SI_Validation_Dist}
\end{figure}

% 日本語コメント：traveling wave の(d),(e)でGor'kov と Baresch がほぼ一致し，いずれも A–A より「やや」小さいことを明示。主眼は(f)の散乱力。
Figure~\ref{fig:SI_Validation_Dist} shows the spatial distribution of the acoustic radiation force at the particle size used in the experiments \((a=0.75~\mathrm{mm},\, \frac{a}{\lambda}= 0.088)\). In the traveling wave case [Fig.~\ref{fig:SI_Validation_Dist}(d)–\ref{fig:SI_Validation_Dist}(f)], the Gor'kov potential and the Baresch formulation behave similarly in Fig.~\ref{fig:SI_Validation_Dist}(d) and Fig.~\ref{fig:SI_Validation_Dist}(e) and give values that are slightly below the Andersson and Ahrens result. The key feature is Fig.~\ref{fig:SI_Validation_Dist}(f): the Gor'kov potential yields an axial force close to zero and is inadequate, whereas the Baresch formulation gives a value that is slightly larger than the Andersson and Ahrens result. This indicates that the Baresch formulation successfully captures the scattering force to the axial force.

\newpage
% =================================================================================

% =================================================================================
% 日本語コメント：本節の導入。放射力に加えて音響流を力モデルに組み込み，Stone らの枠組みに従って流れ場を計算し，その流れから粒子への流体力（drag, lift）を評価することを述べる。
\section{Acoustic Streaming}
This section incorporates acoustic streaming into the force model in addition to acoustic radiation forces. We compute the streaming flow field using the framework established and experimentally validated by Ref.~\cite{stone2025experimental}. The computed flow field is then used to evaluate the hydrodynamic loads on the particle (drag and lift).

\subsection{Governing Equations for Acoustic Streaming}
The streaming field is driven by a time-averaged body force arising from attenuation of acoustic energy in the bulk of the fluid (thermoviscous or atmospheric) \cite{stone2025characterising}:
\begin{equation}
    \frac{\boldsymbol{F}_s}{dV} = \frac{2\alpha\,\boldsymbol{I}}{c_0}.
\end{equation}
where \(\alpha\) is the acoustic attenuation coefficient, and \(\boldsymbol{I}=\tfrac{1}{2}\mathrm{Re}(p\,\boldsymbol{v}^{*})\) is the acoustic intensity vector.

% 日本語コメント：減衰係数の計算に用いたモデル、仮定した環境条件（温度・湿度）、および再現に必要な全ての計算式を詳述する。
\subsection{Attenuation Coefficients}
The analysis of attenuation considered effects from two primary physical models: a thermoviscous model (\(\alpha_{\text{therm}}\)) and an atmospheric model (\(\alpha_{\text{atm}}\)). The thermoviscous model is described using the Stokes-Kirchhoff relation:
\begin{equation}
    \alpha_{\text{therm}}=\frac{\omega^{2}}{2\rho_0 c_0^{3}}\left(\frac{4\mu}{3}+\mu_{B}+(\gamma-1)\frac{k_{\text{cond}}}{C_{p}}\right),
\end{equation}
where \(\mu=1.81\times10^{-5}\,\text{Pa}\cdot\text{s}\) is the dynamic viscosity, \(\mu_B=1.09\times10^{-5}\,\text{Pa}\cdot\text{s}\) is the bulk viscosity, \(\gamma=1.4\) is the ratio of specific heats, \(k_{\text{cond}}=0.02577\,\text{W\,m}^{-1}\text{K}^{-1}\) is the thermal conductivity, and \(C_p=1005.4\,\text{J\,kg}^{-1}\text{K}^{-1}\) is the specific heat capacity at constant pressure. 

For the atmospheric attenuation, we adopted the model by Bass et al.~\cite{bass1995atmospheric}, assuming a temperature of \(20^{\circ}\text{C}\), a relative humidity of 50\%, and an absolute pressure \(p_{\text{abs}}\) of 101325 Pa:
\begin{equation}
    \alpha_{\text{atm}} = f^{2}\left(B_{1}\cdot\frac{f_{rN}}{f_{rN}^{2}+f^{2}}+B_{2}\cdot\frac{f_{rO}}{f_{rO}^{2}+f^{2}}+B_{3}\right),
\end{equation}
where \(f\)=40 kHz is the frequency. The relaxation frequencies for nitrogen (\(f_{rN}\)) and oxygen (\(f_{rO}\)) are computed as:
\begin{align}
    f_{rN} &= \left(\frac{p_{\text{abs}}}{p_{\text{ref}}}\right)\cdot\left(\frac{T_{\text{ref}}}{T_k}\right)^{\frac{1}{2}}\left[9+28000 \cdot \phi_A \cdot \exp\left(-4.17\cdot\left(\left(\frac{T_{\text{ref}}}{T_k}\right)^{\frac{1}{3}}-1\right)\right)\right], \\
    f_{rO} &= \left(\frac{p_{\text{abs}}}{p_{\text{ref}}}\right)\cdot\left[24+\frac{4.04\times10^{6} \cdot \phi_A \cdot (0.02+100\phi_A)}{0.391+100\phi_A}\right].
\end{align}
The coefficients \(B_1, B_2, B_3\) are given by:
\begin{align}
    B_1 &= 0.1068 \left(\frac{T_{\text{ref}}}{T_k}\right)^{2.5} \exp\left(-\frac{3352}{T_k}\right), \\
    B_2 &= 0.01275 \left(\frac{T_{\text{ref}}}{T_k}\right)^{2.5} \exp\left(-\frac{2239.1}{T_k}\right), \\
    B_3 &= 1.84\times10^{-11} \sqrt{\frac{T_k}{T_{\text{ref}}}} \frac{p_{\text{ref}}}{p_{\text{abs}}}.
\end{align}
In these equations, \(T_k = T + 273.16\) is the absolute temperature in Kelvin, \(T_{\text{ref}}=293.15\,\text{K}\) is the reference temperature, and \(p_{\text{ref}}=1\,\text{atm}\) is the reference pressure. The humidity parameter, \(\phi_A\), is calculated from the relative humidity \(\phi\) and the saturation vapor pressure \(p_{\text{sat}}\) as \(\phi_A = \phi\cdot\left(\frac{p_{\text{sat}}}{p_{\text{abs}}}\right)\), where \(p_{\text{sat}}\) is given by
\begin{equation}
    p_{\text{sat}} = p_{\text{ref}}\cdot 10^{4.6151 - 6.8346\cdot\left( \frac{273.16}{T_k} \right)^{1.261}}.
\end{equation}
As demonstrated in Ref.~\cite{stone2025experimental, stone2025characterising}, these two models provide a predicted range for the streaming effects that experimental streaming velocities lie between the predictions of the two models.

\newpage
\subsection{Numerical Implementation}
\begin{figure}[h]
    \centering
    \includegraphics[width=0.8\textwidth]{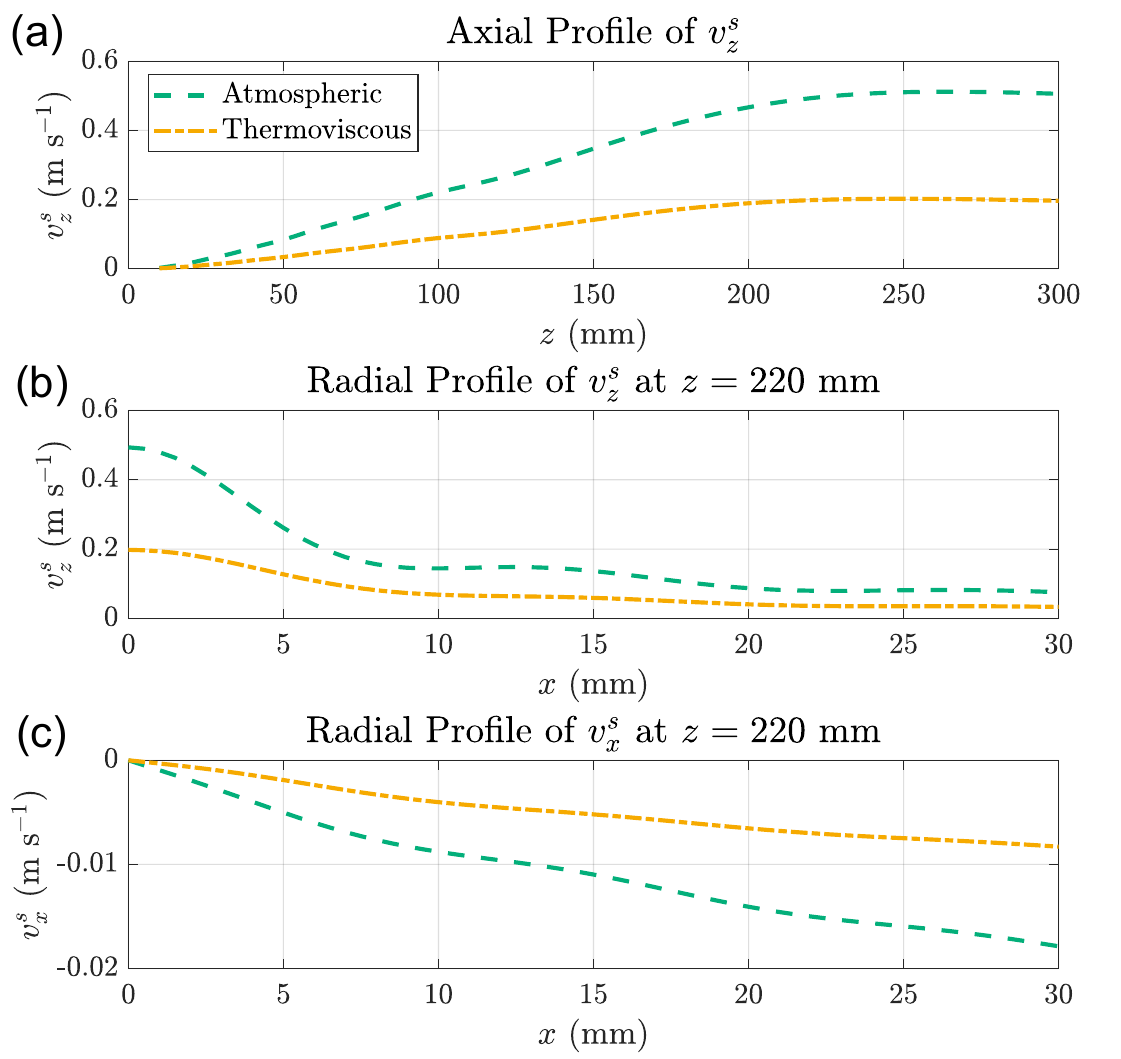}
    \caption{Acoustic streaming velocity field calculated using COMSOL Multiphysics for a zero-order Bessel beam with a cone angle of \(\beta = 20^\circ\) and an input voltage of \(V_{\text{in}} = 8\,\text{V}\). (a) Axial profile of the streaming axial component $v^{s}_{z}$ along the centerline ($x=0$). (b) Radial profile of the streaming axial component $v^{s}_{z}$ at $z=220\,\text{mm}$. (c) Radial profile of the streaming radial component $v^{s}_{x}$ at $z=220\,\text{mm}$.}
    \label{fig:SI_stream_vel}
\end{figure}

To solve for the steady-state flow field, the body force, which was pre-calculated in MATLAB, was first exported as a CSV file. This force field was then imported into a CFD simulation in COMSOL Multiphysics 6.2 and applied as a volume force via interpolation. Our simulation used a two-dimensional axisymmetric model under the assumption of laminar flow. The domain was a \(200 \times 400\)~mm rectangle with a no-slip wall condition at the PAT boundary (\(z=0\)) and pressure outlets at the other boundaries. The resulting steady-state velocity field, calculated for a Bessel beam with $\beta = 20^\circ$ and $V_\text{in}=8\,\text{V}$, is shown in Fig.~\ref{fig:SI_stream_vel}. The simulation reveals a strong upward flow along the central axis of the beam, reaching speeds of 0.198 m\,s\(^{-1}\) (thermoviscous) to 0.493 m\,s\(^{-1}\) (atmospheric) with lateral inflow from the surroundings that converges toward the axis and feeds the upward jet.
% 日本語コメント：具体的な数値計算の実装方法について述べる。COMSOL Multiphysicsを用いたこと、計算領域や境界条件の設定など、再現性に必要な情報を提供する。

\newpage
% 日本語コメント：音響流が粒子に及ぼす抗力の計算方法を詳述する。抗力係数の算出に用いたレイノルズ数依存の経験式も明記する。
\subsection{Drag Force Calculation}
The drag force, \(\boldsymbol{F}_D\), exerted by the streaming flow on the spherical stational particle was calculated from the local streaming velocity field, \(\boldsymbol{v}^{\text{s}}\). The drag force vector is given by:
\begin{equation}
    \boldsymbol{F}_D = - \frac{1}{2} C_d \rho_0 (\pi a^2) |\boldsymbol{v}^{\text{s}}| \boldsymbol{v}^{\text{s}},
\end{equation}
where \(C_d\) is the drag coefficient. The drag coefficient is a function of the (particle) Reynolds number, Re, which is defined as:
\begin{equation}
    \text{Re} = \frac{\rho_0 |\boldsymbol{v}^{\text{s}}| (2a)}{\mu}.
\end{equation}
Since the maximum Reynolds number in our simulations was approximately 50, we employed an empirical correlation applicable for Re \(\leq\) 100~\cite{massey1998mechanics, flemmer1986drag}:
\begin{equation}
    C_d = \frac{24}{\text{Re}}\sqrt{1 + \frac{3}{16}\text{Re}}.
\end{equation}
This calculated drag force vector was then added to the acoustic radiation force and gravity to determine the final equilibrium position of the particle, as shown in Fig.~1(e) of the main text.

%---------------------------------------------------------------------------
\subsection{Lift Force Calculation and Estimation}
\label{sec:SM_lift}
We evaluated the influence of the restoring transverse lift force acting on a stationary particle arising from the shear flow [Fig.~\ref{fig:SI_stream_vel}(b)] established by the acoustic streaming. This flow is characterized by a (particle) Reynolds number (\(\text{Re}\)) and a dimensionless shear parameter (\(\alpha_s\)). The shear parameter are defined as:
\begin{align}
    \alpha_s &= \frac{|\frac{\partial v_z^\text{s}}{ \partial x}| a}{|\boldsymbol{v}^{\text{s}}|}.
\end{align}
where \(v_z^\text{s}\) is the axial component of the streaming velocity \(\boldsymbol{v}^{\text{s}}\).

Under these conditions of a finite Reynolds number, the classical Saffman lift force~\cite{saffman1965lift}, which is strictly valid only for \(\text{Re} \ll 1\), is not directly applicable. Therefore, we employ the model developed by Mei~\cite{mei1992approximate}. This model provides a correction to the Saffman theory, applicable for \(\text{Re}\) and \(\alpha\) within the ranges \(0.1 \le \text{Re} \le 100\) and \(0.005 \le \alpha_s \le 0.4\), respectively.

The classical Saffman lift force, \(F_{L(\text{Saff})}\), which forms the basis for the correction, is given by:
\begin{equation}
    F_{L(\text{Saff})} = 6.46 \rho_0 \sqrt{\nu} a^2 v_z^\text{s} \sqrt{\left|\frac{\partial v_z^\text{s}}{\partial x}\right|} \text{sgn}\left(\frac{\partial v_z^\text{s}}{\partial x}\right),
    \label{eq:SaffmanForce}
\end{equation}
where \(\nu = \frac{\mu}{\rho_0}\) is the kinematic viscosity of the fluid. Mei's model applies the following correction factor, \(C\), to this base force (\(F_L = F_{L(\text{Saff})} \times C\)):
\begin{equation}
C =
    \begin{cases}
        (1 - 0.3314\sqrt{\alpha_s})\exp\left(-\frac{\text{Re}}{10}\right) + 0.3314\sqrt{\alpha_s}, & \text{Re} \le 40 \\
        0.0524\sqrt{\alpha_s \text{Re}}, & \text{Re} > 40.
    \end{cases}
    \label{eq:MeiForce}
\end{equation}

Based on this model, the peak lift force is $|F_L|_{\max} = 3.53\times10^{-9}\ \text{N}$ and $7.08\times10^{-9}\ \text{N}$ for the thermoviscous and atmospheric models, respectively. This force acts as a restoring force directed towards the beam axis. At these locations, the local parameters are $(\mathrm{Re},\alpha_s)=(12.9,\,0.119)$ for the thermoviscous case and $(32.0,\,0.150)$ for the atmospheric case, both well within the model’s validity ranges. These lift forces are two orders of magnitude smaller than the primary transverse acoustic radiation force (on the order of $10^{-7}\ \text{N}$; see Fig.~1(c) in the main text), indicating that shear-induced aerodynamic lift is not a dominant contributor to the transverse stability of the Bessel beam. Consistently, stable levitation is also observed for non-spherical
particles.
%=================================================================================

% 日本語コメント：このセクションでは、実験的に観測された粒子の浮上位置データとその分析手法、およびその他の浮上実例について詳述します。
\newpage
\section{Experimental Levitation Data and Analysis}
% 日本語コメント：このサブセクションでは、時系列データから安定な浮上位置を特定するための、実験、信号処理、およびクラスタリング手法について、一連の流れとして記述します。
\subsection{Experimental Analysis of Discrete Levitation Positions}
\begin{figure}[h]
    \centering
    \includegraphics[width=0.95\textwidth]{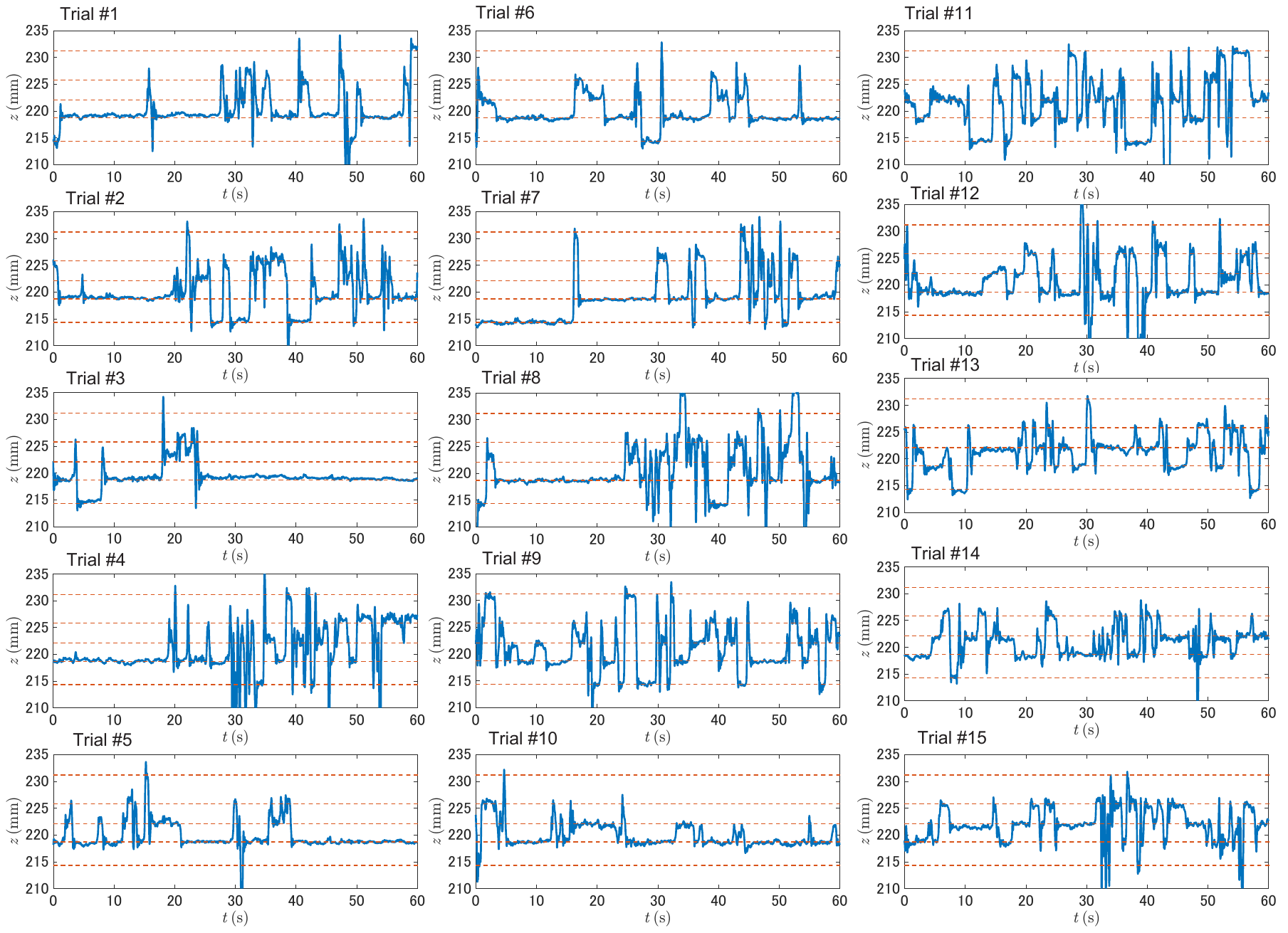} % Provide the correct path to your figure file
    \caption{Time-series data of the axial levitation position \( z \) of an EPS sphere (blue line) from a representative trial. The particle's trajectory shows levitation characterized by spontaneous transitions between discrete heights. The red dashed lines indicate the levitation heights (214.3, 218.7, 222.1, 225.8, and 231.2 mm) identified by applying a clustering analysis to the data from all 15 trials.}
    \label{fig:SI_t_vs_z}
\end{figure}

Fifteen independent 60 s levitation trials were conducted using a zero-order Bessel beam with $\beta = 20^{\circ}$ and $V_\text{in}=8\,\text{V}$. In each trial, an EPS sphere ($a=0.75\,\text{mm}$) was levitated, and its time-series position data was recorded. The data revealed spontaneous transitions between several discrete levitation positions, as shown in Fig.~\ref{fig:SI_t_vs_z}. To quantitatively identify these stable heights from the time-series data, transient fluctuations were first removed using change detection algorithms combined with statistical methods involving moving averages and standard deviations. Subsequently, k-means clustering was applied to the consolidated data from all 15 trials. The optimal cluster number was determined using the silhouette method, and clusters with centers within 1 mm of each other were consolidated. This analysis revealed five distinct stable levitation heights, with cluster centers identified at 214.3, 218.7, 222.1, 225.8, and 231.2 mm.
% 日本語コメント：この図は、単一粒子が複数の離産的な高さ間を遷移する様子を示す時系列データです。キャプションで、クラスタリングによって特定された安定浮上高さを具体的に示します。

\newpage
% Japanese comment: Report examples of non-spherical samples and operating conditions (\beta, V_in).
\subsection{Levitation of Non-Spherical Objects}
\begin{figure}[t]
    \centering
    \includegraphics[width=0.40\textwidth]{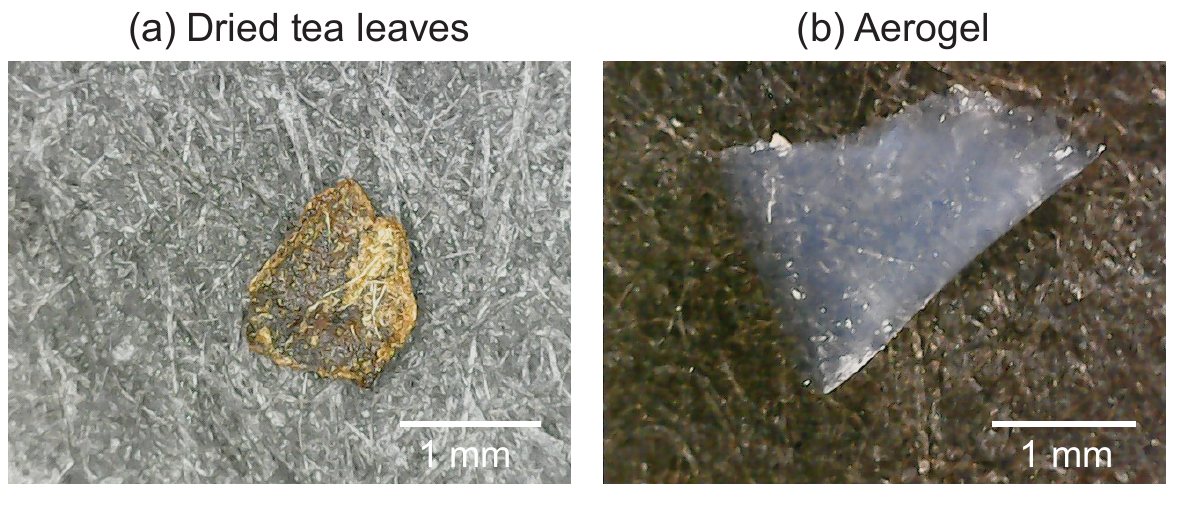}
    \caption{Photographs of non-spherical objects levitated in the experiments. (a) A dried tea leaf, levitated at \(V_\text{in}=16\,\mathrm{V}\) (Movie~S3). (b) A piece of silica aerogel, levitated at \(V_\text{in}=8\,\mathrm{V}\) (Movie~S4). Both experiments used a zero-order Bessel beam with \(\beta=20^{\circ}\).}
    \label{fig:SI_other_objects}
\end{figure}

The levitation capability of the Bessel beam is not limited to spherical particles. Levitation was also demonstrated for various non-spherical objects. For instance, a dried tea leaf [Fig.~\ref{fig:SI_other_objects}(a), Movie~S3] and a piece of silica aerogel [Fig.~\ref{fig:SI_other_objects}(b), Movie~S4] were levitated using a zero-order Bessel beam with \(\beta=20^{\circ}\). Input voltages were \(V_\text{in}=16\,\mathrm{V}\) for the tea leaf and \(V_\text{in}=8\,\mathrm{V}\) for the aerogel. Accordingly, stability mechanisms that rely on spherical symmetry, such as sphere-specific aerodynamics or symmetry-dependent acoustic scattering, can be excluded as primary contributors in our method.

% Japanese comment: Summarize what the figure shows (irregular objects and experimental conditions).

% 日本語コメント：このセクションでは、水平および垂直方向の粒子操作実験における、粒子の軌跡の定量的データを示す。
%\newpage
\section{Analysis of Particle Translation}
% 日本語コメント：この図は、水平・垂直操作時の粒子の追従性を示す時系列データです。キャプションを簡潔にし、各パネルの内容を説明します。
\begin{figure}[b]
    \centering
    \includegraphics[width=0.60\textwidth]{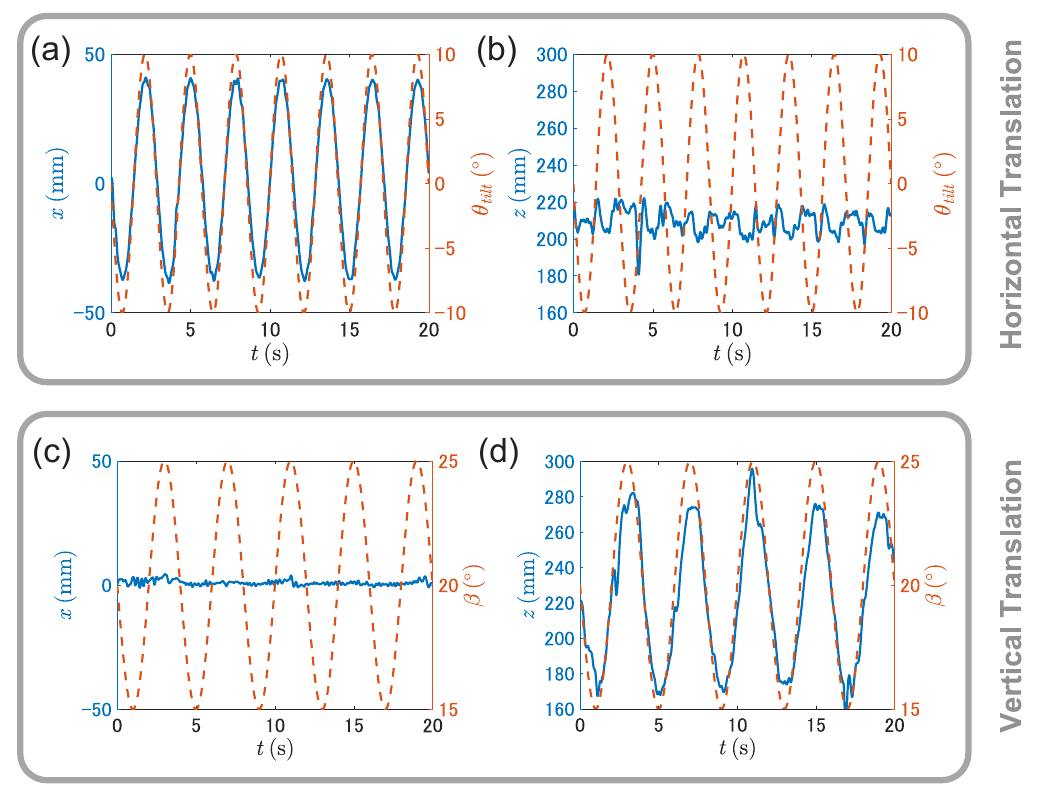}
    \caption{Time-series response of the particle to commanded beam modulations. In all panels, blue (left axis) shows the measured particle coordinate and red dashed (right axis) shows the command signal (tilt angle \(\theta_{\text{tilt}}\) or cone angle \(\beta\)). Horizontal translation (see also Figs.~2(a)-2(c) in the main text): (a) \(x(t)\) with commanded \(\theta_{\text{tilt}}(t)\); (b) concurrent \(z(t)\). Vertical translation (see also Figs.~2(d)-2(f) in the main text): (c) \(x(t)\) with commanded \(\beta(t)\); (d) concurrent \(z(t)\).}
    \label{fig:SI_manipulate_tracking}
\end{figure}

% Japanese comment: Summarize time response and tracking; avoid subjective wording and follow panel order.
Figure~\ref{fig:SI_manipulate_tracking} shows time-series traces of the particle positions in response to sinusoidal modulation of the beam tilt (Fig.~2(a)-2(c) in the main text) and cone angle (Fig.~2(d)-2(f) in the main text). Under tilt modulation, the horizontal coordinate \(x\) tracked the modulated tilt angle \(\theta_{\text{tilt}}\) [Fig.~\ref{fig:SI_manipulate_tracking}(a)], while the axial coordinate \(z\) exhibited small fluctuations [Fig.~\ref{fig:SI_manipulate_tracking}(b)]. Under cone-angle modulation, the horizontal coordinate remained near the beam axis [Fig.~\ref{fig:SI_manipulate_tracking}(c)], and the axial coordinate \(z\) tracked the modulated cone angle \(\beta\) [Fig.~\ref{fig:SI_manipulate_tracking}(d)].

% 日本語コメント：本セクション全体の導入。作動範囲の実験的決定手法の概要、共通の実験条件、データ取得・処理方法について記述する。
%\newpage
\section{Experimental Determination of the Working Range}
The working ranges of the Bessel beam and the twin trap, presented in Fig.~2(g) of the main text, were determined using the following experimental procedure. The working range was defined by the spatial limits of particle manipulation. For all trials, an EPS sphere with a radius of $a = 0.75\,\text{mm}$ was used. The centroid coordinates were extracted via an image processing pipeline (Python 3.11, OpenCV) that included background subtraction, Gaussian blurring, thresholding, and connected components analysis. The system was calibrated against a reference object, yielding conversion factors of $0.16\,\text{mm pixel}^{-1}$ for the twin trap and $0.61\,\text{mm pixel}^{-1}$ for the Bessel beam. To ensure reliability, the entire procedure was repeated five times for each trap configuration, and the results were averaged.

% 日本語コメント：ツイントラップの作動範囲を特定した具体的な実験手順を記述する。
The working range of the twin trap was mapped by translating its control point from an initial position at $z = 20\,\text{mm}$. The control point was moved outward along predefined radial trajectories at angles of $-90^\circ$, $0^\circ$, $30^\circ$, $60^\circ$, and $90^\circ$ relative to the array's normal. Along each trajectory, the displacement was increased in $1\,\text{mm}$ increments. The boundary of the working range was defined as the position where the trap failed and the particle was ejected.

% 日本語コメント：ベッセルビームの作動範囲を特定した、より複雑な二段階の実験手順を記述する。
The working range of the Bessel beam was mapped using a two-step procedure. First, to define the axial limits, we determined the operational range of the cone angle, $\beta$, that supports stable levitation. Starting from a stable trap at $\beta = 20^\circ$, the angle was incrementally adjusted in $0.1^\circ$ steps until levitation failed, yielding a viable range from $\beta_{\min} = 11.4^\circ$ to $\beta_{\max} = 27.8^\circ$. Second, this angular range was discretized into five equidistant setpoints. At each setpoint, the beam was progressively tilted in $0.1^\circ$ increments until the particle was ejected. The particle's coordinates at the moment of trap failure were recorded, and the envelope of these boundary points defined the working range of the Bessel beam.

\newpage
% 日本語コメント：このセクションでは、本文で議論された横方向の復元力とトラップ安定性について、数値計算を用いて物理的な裏付けを提供します。
\section{Analysis of Transverse Forces and Trap Stability}

% 日本語コメント：このサブセクションでは、Gor'kovポテンシャルの圧力項と速度項を分離することで、ベッセルビームが高圧領域で復元力を持てるのか、その物理的メカニズムを分析します。
\subsection{Decomposition of the Transverse Restoring Force}
\begin{figure}[h]
    \centering
    \includegraphics[width=0.85\textwidth]{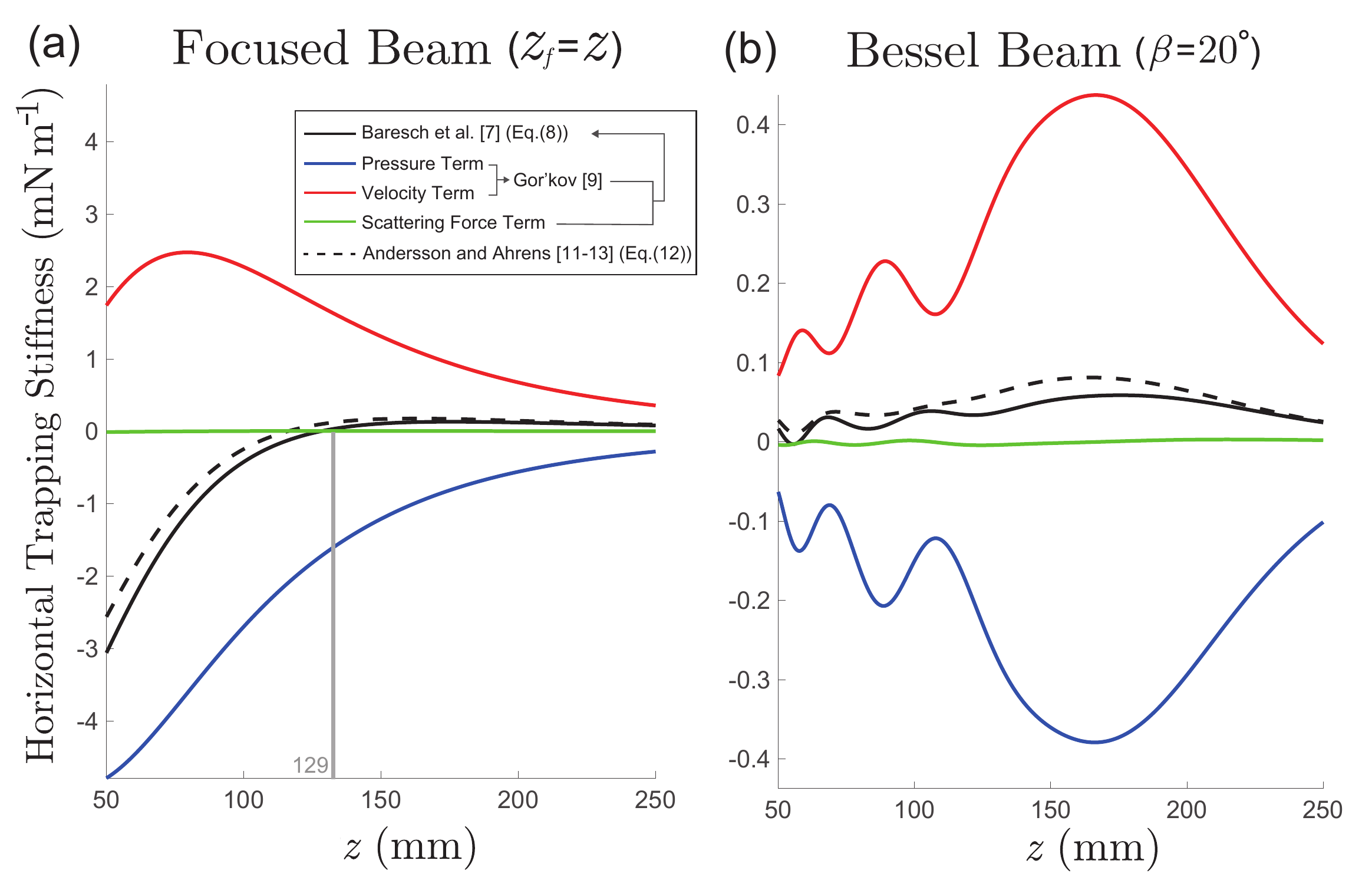}
    % 日本語コメント：箇条書き調を避け，図が何を示すかと前提（zf=z）だけを簡潔に文章で記述する。
    \caption{Decomposition of the horizontal trapping stiffness (\(-\pdv{F_x}{x}\)) for (a) a conventional focused beam and (b) a zero-order Bessel beam (\(\beta=20^{\circ}\)). The solid black curve shows the total stiffness computed using Baresch formulation [Eq.~\eqref{eq:BareschForce}]. The blue and red curves show, respectively, the contributions from the pressure-gradient and velocity-gradient terms, and the green curve shows the small scattering force contribution. For comparison, the black dashed curve shows the stiffness computed with the expansion in spherical harmonics due to Andersson and Ahrens [Eq.~\eqref{eq:Fx}]. In panel (a), the focal position is set equal to the axial evaluation point, i.e., \(z_f = z\).}
    \label{fig:SI_gorkov_components}
\end{figure}

To elucidate the origin of the horizontal trapping stiffness, $-\pdv{F_x}{x}$, we decompose the total force in Eq.~\eqref{eq:BareschForce} into its constituent terms for both a conventional focused beam and a zero-order Bessel beam [Fig.~\ref{fig:SI_gorkov_components}]. For the focused beam [Fig.~\ref{fig:SI_gorkov_components}(a)], the pressure-gradient term (first term in Eq.~\eqref{eq:BareschForce}) yields a repulsive contribution (negative stiffness), whereas the velocity-gradient term (second term) yields an attractive one (positive stiffness). The total stiffness becomes positive only for $z \gtrsim 129\,\mathrm{mm}$, where the attractive velocity contribution exceeds the repulsive pressure contribution. For the zero-order Bessel beam with $\beta=20^{\circ}$ [Fig.~\ref{fig:SI_gorkov_components}(b)], the attractive velocity-gradient term dominates its weaker repulsive counterpart over the plotted range. The contribution of the scattering force terms (remaining terms in Eq.~\eqref{eq:BareschForce}) to the horizontal stiffness is negligible in both configurations.

% \newpage
% % Japanese comment: Use \sim for approximations; unify units, degree symbol, and terminology across the manuscript.
% \subsection{Discrepancy between Predicted Stiffness and Observed Stability}
% The numerically computed horizontal trapping stiffness is inconsistent with the stability observed in our experiments. As shown in Fig.~\ref{fig:SI_stiffness_comparison}, the calculation yields a higher peak stiffness for a conventional focused beam (about \(0.17\,\mathrm{mN\,m^{-1}}\) at \(z \sim 160\,\mathrm{mm}\)) than for the Bessel beam, which would suggest stronger transverse confinement for the focused case. In contrast, under comparable conditions sustained levitation was obtained only with the Bessel beam. This discrepancy indicates that the observed stability is not fully captured by the present stiffness metric and warrants further investigation.

% \begin{figure}[h]
%     \centering
%     \includegraphics[width=0.85\textwidth]{figs/SI_trapping_stiffness.pdf}
%     \caption{Comparison of the horizontal trapping stiffness, \(-\pdv{F_x}{x}\), as a function of axial position \(z\), computed using the Andersson and Ahrens method [Eqs.~\eqref{eq:Fx}–\eqref{eq:Fz}]. For the focused beam (red solid), the focal position is set equal to the axial evaluation point, \(z_f = z\). For zero-order Bessel beams (blue), cone angles are \(\beta=15^{\circ}\) (dashed), \(20^{\circ}\) (solid), and \(25^{\circ}\) (dotted). A positive value indicates a restoring stiffness toward the beam axis.}
%     \label{fig:SI_stiffness_comparison}
% \end{figure}

\newpage
% Japanese comment: Keep terminology, units, and derivative notation consistent with the rest of the manuscript.
\subsection{Comparison of Levitation-Capable Regions}

\begin{figure}[h]
    \centering
    \includegraphics[width=0.95\textwidth]{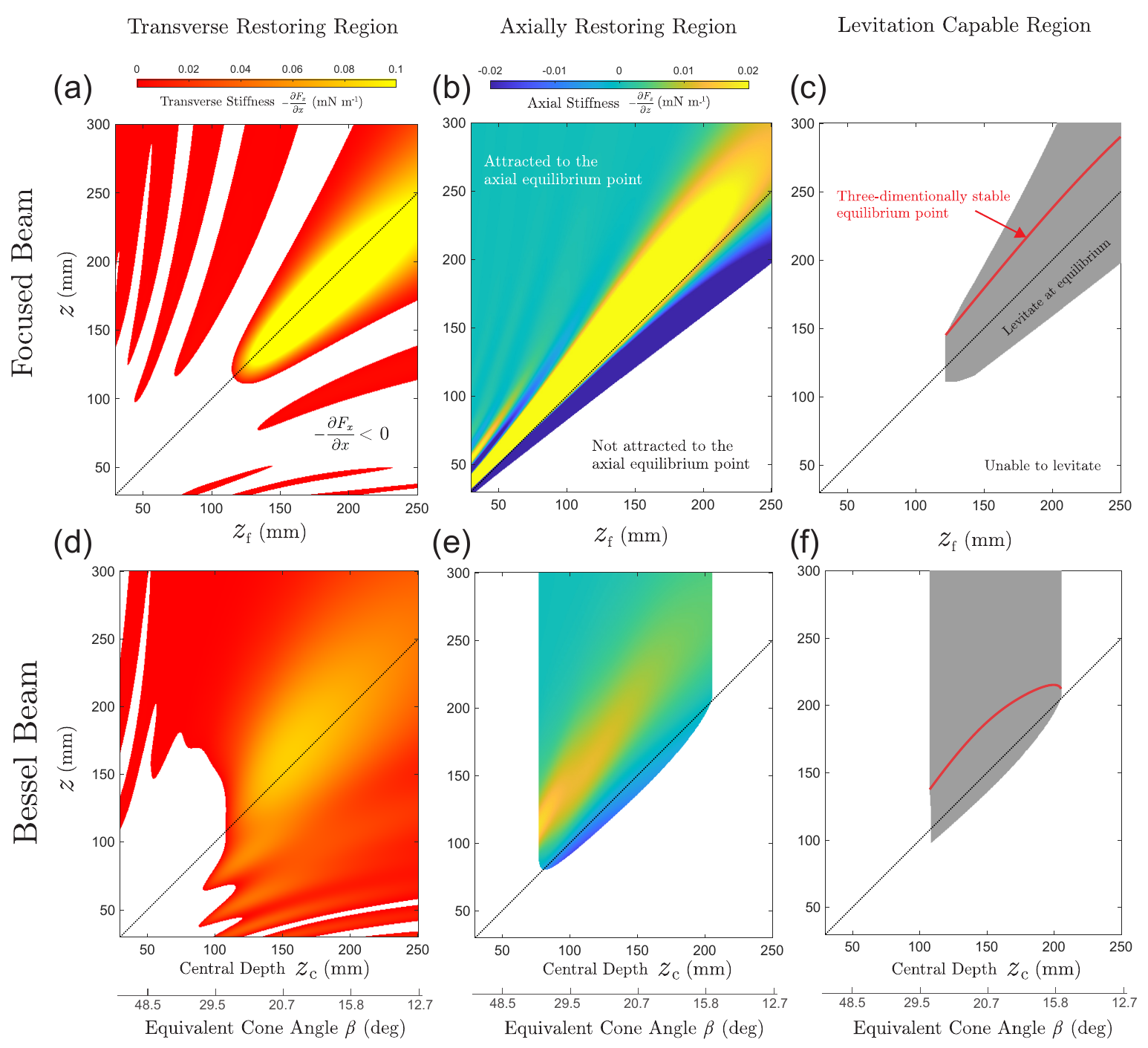}
    \caption{Stability maps for a conventional focused beam and a zero-order Bessel beam. All plots show forces computed on the beam axis for an EPS sphere (\(a=0.75\,\mathrm{mm}\)) at an input voltage of \(V_{\text{in}}=8\,\mathrm{V}\). Acoustic radiation force is computed using the Andersson and Ahrens method [Eqs.~\eqref{eq:Fx}–\eqref{eq:Fz}]. The vertical axis is the on-axis position \(z\), while the horizontal axis is the focal length \(z_f\) (focused beam) or the central depth \(z_c=\frac{\sqrt{2}A}{4\tan{\beta}}\) (Bessel beam of a finite aperture). (a),(d) Transverse restoring regions, where transverse stiffness is positive. (b),(e) Axial restoring regions, where the axial force is directed toward a stable local equilibrium. (c),(f) The resulting levitation-capable region (gray area). A point is included only if it satisfies both transverse and axial restoring conditions, and if a stable 3D equilibrium point (red curve) exists within the same contiguous stable region for that beam configuration (\(z_f\) or \(z_c\)).}
    \label{fig:DW_stiffness_mask}
\end{figure}

To systematically compare the stability of the conventional focused beam and the zero-order Bessel beam, we computed maps of their levitation-capable regions [Fig.~\ref{fig:DW_stiffness_mask}]. For the focused beam, the levitation-capable region [Fig.~\ref{fig:DW_stiffness_mask}(c)] is bounded by an upper limit because the transverse restoring region vanishes above this axial position [Fig.~\ref{fig:DW_stiffness_mask}(a)], which makes the trap susceptible to vertical perturbations. In contrast, the zero-order Bessel beam sustains a transverse restoring region at higher axial positions [Fig.~\ref{fig:DW_stiffness_mask}(d)], yielding a more extensive levitation-capable region [Fig.~\ref{fig:DW_stiffness_mask}(f)].

\section{Reflection from the ceiling}
%【段落の役割】実験で観測された複数の離散的な浮上高さと理論予測との矛盾を提示し、その原因を天井からの反射と仮定して数値モデルで検証し、微弱な反射でさえも浮上ダイナミクスに大きな影響を及ぼすため、今後の設計では環境反射の考慮が不可欠であると結論づける。
Our numerical model predicts a single equilibrium point (Fig.~1(e) in the main text). However, we observed the particle transitioning between multiple discrete levitation heights in experiments [Fig.~\ref{fig:SI_t_vs_z}]. We attribute this discrepancy to weak reflections from the ceiling, located approximately 1.55 m above the PAT.

\begin{figure}[h]
    \centering
    \includegraphics[width=0.95\textwidth]{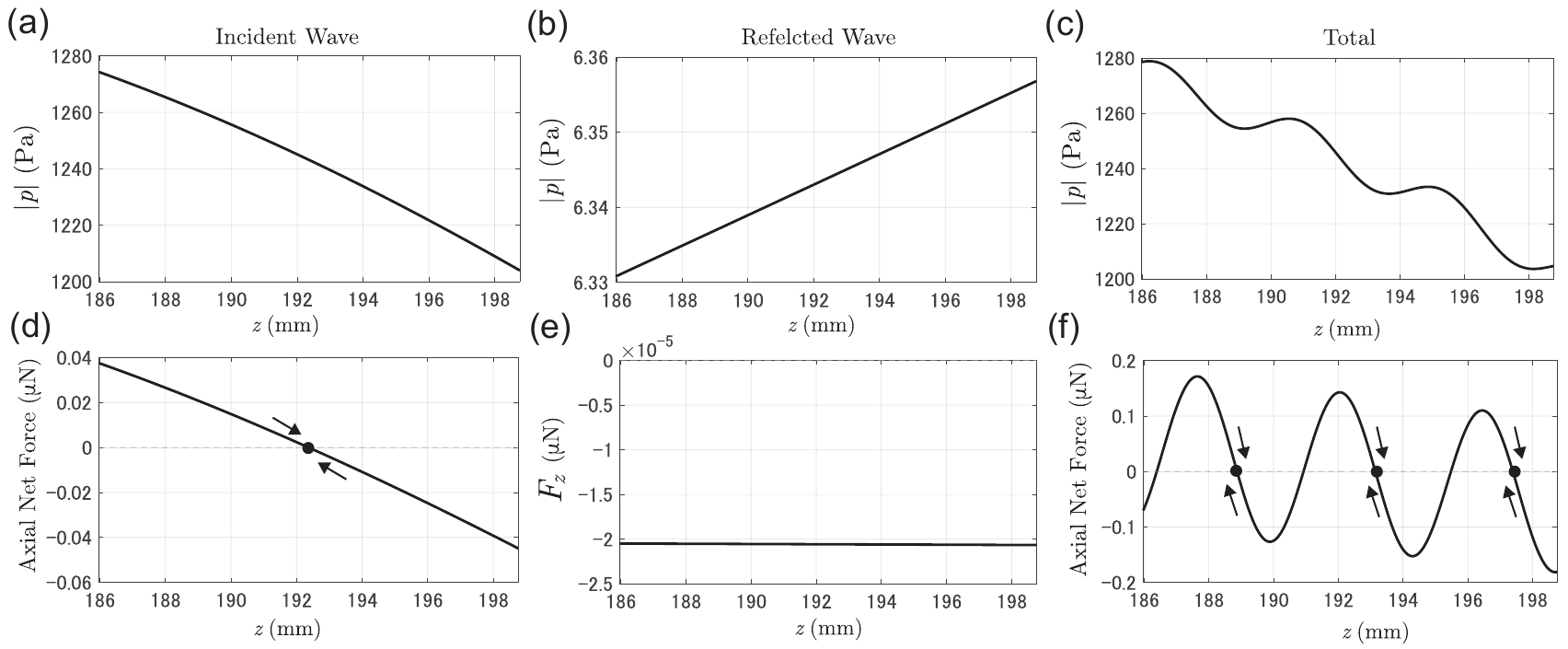}
    % 日本語コメント：箇条書き調を避け，図が何を示すかと前提（zf=z）だけを簡潔に文章で記述する。
    \caption{Numerical calculation of the effect of a worst-case ceiling reflection on the axial acoustic pressure and net axial force for a zero-order Bessel beam ($\beta = 20^\circ$, $V_\text{in}=8\,\text{V}$) on an EPS sphere ($a=0.75\,\text{mm}$), with forces calculated using the Andersson and Ahrens method [Eqs.~(\ref{eq:Fx})–(\ref{eq:Fz})]. The plots compare (a, d) the incident wave from an upward-facing PAT at $z_i=0$ m, (b, e) the reflected wave from the ceiling, and (c, f) their superposition. (a)–(c) Axial pressure profile. The net axial force (radiation force and gravity) is shown in (d) and (f), while (e) shows only the radiation force $F_z$. The worst-case reflection is modeled from a downward-facing virtual PAT at $z_i=3.1\,\text{m}$ (twice the ceiling height of 1.55 m), which assumes perfect reflection and neglects factors such as absorption in the medium.}
    \label{fig:reflection}
\end{figure}

We constructed a simple model based on Huygens' principle model [Eq.~\eqref{eq:sm_pressure}], with two sets of PAT array facing each other. The upper array (positioned 3.1 m away from the origin) virtually emulates a perfectly reflected wave. The reflected wave has an amplitude of $\sim6$ Pa ($<$1\% of the incident beam pressure). While constructing quantitatively matching model is challenging (due to more complex reflections, absorptions in real environment); this simple model qualitatively recovers multiple local equilibria [Fig.~\ref{fig:reflection}(f)] observed in Fig.~\ref{fig:SI_t_vs_z}].

% 日本語コメント：このセクションでは、本文および補足資料で言及されている全ての動画について、その内容を説明するキャプションを記載します。
\newpage
\section*{Movie Captions}

\textbf{Movie S1: Instability of a Conventional Focused Beam}
\newline
An expanded polystyrene (EPS) sphere (radius \(a=0.75\) mm) is ejected from a conventional focused beam (\(z_f = 180\) mm). Despite an initial capture, the particle is expelled after approximately 3 seconds, demonstrating the trap's inherent instability.
\vspace{0.3cm}

\textbf{Movie S2: Stable Levitation with Spontaneous Transitions}
\newline
Stable levitation of an EPS sphere (\(a=0.75\) mm) within a zero-order Bessel beam (\(\beta=20^\circ\), \(V_\text{in}=8\,\text{V}\)) is shown. The video first presents a representative trial (\#3) with occasional spontaneous transitions between discrete levitation heights, followed by another trial (\#9) exhibiting more frequent transitions.
\vspace{0.3cm}

\textbf{Movie S3: Levitation of a Dried Tea Leaf}
\newline
An non-spherical shaped dried tea leaf (approximate diameter, 0.8 mm) is maintained in a levitated state for one minute within a zero-order Bessel beam (\(\beta=20^\circ\), \(V_\text{in}=16\,\text{V}\)). Although the particle exhibits large vertical oscillations, its levitated state is sustained. The footage begins with a wide view of the experimental setup before zooming in on the levitated leaf.
\vspace{0.3cm}

\textbf{Movie S4: Levitation of Silica Aerogel}
\newline
An non-spherical shaped flat piece of silica aerogel (major axis approx. 2.3 mm) is maintained in a levitated state using a zero-order Bessel beam (\(\beta=20^\circ\), \(V_\text{in}=8\,\text{V}\)). The footage starts with a wide view before zooming in. Compared to the tea leaf (Movie S3), the levitation is more stable with smaller oscillations.
\vspace{0.3cm}

\textbf{Movie S5: Levitation of Potato Starch Powder}
\newline
A disk-shaped object (approx. 1.3 mm diameter) formed from coalesced potato starch powder is maintained in a levitated state (\(\beta=20^\circ\), \(V_\text{in}=15\,\text{V}\)), though it exhibits vertical oscillations. Both top and side views are provided.
\vspace{0.3cm}

\textbf{Movie S6: Horizontal Translation}
\newline
Horizontal translation of an EPS sphere (\(a=0.75\) mm) is demonstrated. The particle is manipulated by tilting the Bessel beam, with the tilt angle \(\theta_{\text{tilt}}\) modulated from \(-10^\circ\) to \(+10^\circ\).
\vspace{0.3cm}

\textbf{Movie S7: Vertical Translation}
\newline
Vertical translation of an EPS sphere (\(a=0.75\) mm) is demonstrated. The particle's axial position is controlled by modulating the cone angle \(\beta\) of the Bessel beam from \(15^\circ\) to \(25^\circ\).
\vspace{0.3cm}

\textbf{Movie S8: Parallel Trapping of Two Particles}
\newline
Simultaneous levitation and manipulation of two EPS spheres in parallel are achieved using two separate Bessel beams generated from a single PAT by applying an acoustic Dammann grating phase profile (\(\beta=20^\circ\), \(V_\text{in}=13\,\text{V}\)).
\vspace{0.3cm}

\textbf{Movie S9: Vertically Aligned Trapping of Two Particles}
\newline
Two EPS spheres are levitated simultaneously at discrete axial positions using a bottle beam configuration (\(\beta=21^\circ\) and \(V_\text{in}=9\,\text{V}\)).
\vspace{0.3cm}

\textbf{Movie S10: Levitation over an Obstacle}
\newline
Levitation of an EPS sphere over a physical obstruction is shown. The experimental footage is composited with the sound pressure profile calculated by the boundary element method (BEM).  This demonstrates that the Bessel beam can maintain a stable trap beyond the obstacle, even when the beam path is partially occluded.
\vspace{0.3cm}

\textbf{Movie S11: Analysis of Particle Rotation}
\newline
A marked EPS sphere is shown levitating. The footage is played first at real-time speed and then at 0.1$\cross$ speed to clearly show the lack of rotation, ruling out rotation-dependent forces as a primary stability mechanism.
\vspace{0.3cm}

% \textbf{Movie S12: Simultaneous Levitation at Different Transition Heights}
% \newline
% Two EPS spheres are shown levitating simultaneously at different discrete axial heights. This video corresponds to the quantitative data presented in Fig.~\ref{fig:SI_vertical_align} and demonstrates the existence of multiple stable trapping positions within a single beam.

\newpage
% \bibliographystyle{apsrev4-2}
% \bibliography{mybib} % <-- your main .bib file
%\nocite{*}
\bibliography{mybib}% Produces the bibliography via BibTeX.